\documentclass[aps,prd,preprint,groupedaddress]{revtex4-1}

\usepackage[utf8]{inputenc}     
\usepackage{graphicx}
\usepackage{color}        
     
\newcommand{\be}{\begin{equation}}
\newcommand{\ee}{\end{equation}}
\newcommand{\ba}{\begin{eqnarray}}
\newcommand{\ea}{\end{eqnarray}}
\newcommand{\bi}{\begin{itemize}}
\newcommand{\ei}{\end{itemize}}

\newcommand{\la}{\label}

\newcommand{\FF}{\mathcal{F}}
\newcommand{\non}{{\nonumber}}

\def\a{\alpha}         \def\g{\gamma}      
\def\d{\delta}        
          \def\l{\lambda}     
\def\m{\mu}                     
\def\n{\nu}             
\def\r{\varrho}     \def\s{\sigma}  
\def\t{\tau}

  \def\OO{{\cal O}}

\newcommand\Tr{\mbox{Tr}}

\begin{document}

\preprint{CPT-P002-2014}
\preprint{NIKHEF/2014-016}  
\preprint{WUB/14-05}         

\title{The thermal instanton determinant in compact  form}


\author{Chris. P. Korthals Altes}
\email{altes@cpt.univ-mrs.fr}
\affiliation{Centre Physique Th\'eorique
\footnote{UMR 6207 of CNRS, Universities  Aix-Marseille I and II, 
and of Sud Toulon-Var, and affiliated with FRUMAM.}
Case 907, Campus de Luminy, F-13288 Marseille, France, }
\affiliation{NIKHEF theory group, P.O. Box 41882, 1009 DB Amsterdam, The Netherlands}

\author{Alfonso Sastre}
\email{sastrebruno@uni-wuppertal.de}
\affiliation{Fachbereich C, Bergische Universitaet Wuppertal,, D-42097 Wuppertal, Germany }


\date{\today}  

\begin{abstract}
The thermal instanton determinant for the gauge group $SU(2)$ can be reduced to a form involving two simple functions. 
Only a two dimensional integral has to be done numerically.  
Various boundary conditions are incorporated, in particular an interpolation between bosonic and fermionic 
statistics.
As an example we compute the contribution to the free energy of  ${\cal{N}}=1$ theory.
\end{abstract}


\maketitle

\section{Introduction}

Quantum Chromodynamics (QCD) at  temperatures on the order of the pion mass has remarkable properties 
as shown in RHIC and ALICE/ATLAS experiments \cite{Tannenbaum:2014tea}, and predicted by theoretical papers and lattice simulations~\cite{Szabo:2014iqa}. 

Perturbation theory has only a limited applicability, although the use of effective theories has greatly enhanced its usefulness. 
Semi-classical methods based on instantons~\cite{'tHooft:1976up,'tHooft:1976fv,Belavin:1975fg} have met with qualitative success~ \cite{Shuryak:1987jb}.

The thermal instanton~\cite{Harrington:1978ve} determinant  has stayed a  calculational tour de force~\cite{Gross:1980br} 
since some thirty odd years.  One of the reasons that it was not revisited is the lack
of a phenomenological motive. The contribution to the pressure is quite small due to the typical exponentially suppressed instanton amplitude. More recently, with the advent of
calorons~\cite{Lee:1998bb, Kraan:1998pm} the  question of the instability of the Stefan Boltzmann gas can be studied. 
This is interesting by itself because the caloron~\footnote{We use the word caloron for those periodic instantons which have trivial or  non-trivial holonomy,
i.e. they are spatially asymptoting into a (non-)trivial Polyakov loop. The  case considered here with trivial Polyakov loop is called the thermal instanton, or HS caloron.}
maybe the first semiclassical contribution to do so~\cite{Diakonov:2004jn}. Also recent work on semi-classical methods~\cite{Cherman:2014ofa} spurs a renewed analysis.

In this paper we reexamine the calculation of the fluctuation determinant for the Harrington-Shephard~\cite{Harrington:1978ve} (in the sequel called HS)
thermal instanton in the case of $SU(2)$ gauge theory.   The analysis follows the traditional strategy: we break up the calculation into a part containing the short distance contribution and 
a contribution due to the thermalization. The first  contribution was computed in a series of seminal papers by Brown et al.~\cite{Brown:1977cm,Brown:1977eb,Brown:1978yj}. 
The second, thermal contribution, in a beautiful but intricate, for the isospin 1
mostly numerical analysis by  Gross, Pisarski and Yaffe~\cite{Gross:1980br} (hereafter referred to as GPY), turned out to be
numerically related  to the first contribution in a surprisingly simple way.

Our method translates the first order differential identities~\cite{Brown:1978yj} that served  GPY so nicely   in the isospin ${1\over 2}$ case to similar identities in the isospin $1$ case. We then find through straightforward analytic methods that this isospin 1 thermal contribution actually factorizes.

The first factor is the contribution already computed by Brown et al.~\cite{Brown:1978yj}. 
The second factor consists of combination of a few elementary functions, and contains no reference to the parameters of the HS caloron.
It depends on the boundary conditions (periodic, anti-periodic ....). 
In the periodic case we confirm the numerical analysis of GPY.

The lay out of this paper is as follows. The next section  \ref{sec:general} recalls general aspects underlying our calculation and a 
brief description of the thermal instanton and the propagator of isospin 1/2 and 1 in the field of the thermal instanton. 
 
We then, in section \ref{sec:HS}, discuss our results and discuss their salient properties. The reader only interested in results can skip the other sections.

The next section \ref{sec:determinants} starts with the strategy to obtain our results. This strategy consists of two  stages. 
In the subsections these stages are presented.

Conclusions are contained in section \ref{sec:conclusions}.

In  Appendix \ref{app:fdf} we work out two important differential identities used in section \ref{sec:determinants}. These identities lead, as mentioned above, to huge cancellations and our simple result obtains.
In Appendix \ref{app:sums} the reader can find  useful summation formulas and apply them to his or her favored boundary  condition.

\section{General considerations}\la{sec:general}

Here we will exhibit the main ideas going into the calculation of the thermal case. Throughout we will use the notation of GPY to ease comparison with their work.  

We start with the relevance for the free energy of QCD and how the instanton gas enters into it.

The free energy density of QCD: 

\be
\exp(-V/Tf(T))=\int DA_\m D\psi D\bar\psi \exp\left(-{1\over{g^2}}S(A)\right),
\la{freenergypath}
\ee
has periodic boundary conditions for the vector potentials $A_\m$  and anti-periodic ones for the fermions $\psi$.  
It is expanded in perturbation theory at high temperature $T$ due to asymptotic freedom.
This  expansion can be done around any local minimum of the QCD action $S$.  If the minimum is the 
perturbative one the leading result is due the determinant:
\be
Z_{pert}=\det ((-\partial^{-2}_T)/(-\partial^{-2}))
\la{freeenergypert}
\ee

The four dimensional Dalembertian $\partial^2$ is normalizing the four dimensional Dalembertian $-\partial^2_T$ with periodic time ($t\rightarrow t+1/T$).

\be
\log Z_{pert}=-V/T f_{SB},~ f_{SB}=-{\pi^2\over{15}}T^4
\la{SB}
\ee

\noindent the Stefan-Boltzman free energy for the gauge group $SU(2)$, our main concern.

In this paper we are interested in the fluctuations around the non-trivial minimum given by the self-dual and 
periodic  Harrington-Shephard (HS) instanton, in particular the fluctuation determinant. We denote this single caloron
determinant by $Z_1$ , and $D_1^2$ the spin zero Dalembertian with as background the HS instanton in the adjoint representation of $SU(2)$.  
Then:
\be
Z_{1}=\det(-D_1^{-2}/-\partial_T^{-2})
\la{z1}
\ee
This quantity has to be integrated over the zero-modes (see next section) and yields $\hat Z_1$.

The grand partition function $Z_{gr}$, where we admit an arbitrary number $N_\pm$ of 
HS instanton or anti-instantons in the system becomes:

\ba
Z_{gr}&=&Z_{pert}\sum_{N_=,N_-}{1\over {N_+!}}{1\over{N_-!}}\hat Z_1^{N_++N_-}\non\\
&=&Z_{pert}\exp(2\hat Z_1).
\ea
This expression supposes a very dilute gas of HS instantons, which is valid at very high temperature.
This is due to the thermal screening of the instantons. Surely at asymptotic temperatures the formula makes sense.
So, with Eq. (\ref{SB}), we have:
\be
f_{gr}=f_{SB}-2{T/V}\hat Z_1
\la{totalfreeenergy}
\ee

The fundamental building block for  $\hat Z_1$ is the determinant of a scalar in the  HS background, Eq. (\ref{z1}),  and will be discussed now in more detail. 

Let us denote
the covariant derivative in some representation $R$ of $SU(2)$ by: 
\be
D_R=\partial-i A_R.
\la{covariantderiv}
\ee
\noindent with
\be
A_R=A_R^aT^a_R, ~T^a_{{1\over 2}}=\frac{\t^a}{2},~(T^a_{1})^{ij}=i\epsilon^{iaj}. 
\ee
Consider the determinant of this operator in the spin zero case:
\be
\log\det(-D_R^2).
\ee 
\noindent where $A_R$ is the HS caloron in the representation $R$.

The traditional way to obtain this quantity is to  compute its derivative with respect to one of the parameters
in the HS caloron called $\l$ (see next subsection) and then integrate from $0$ to $\l$:
\ba
\d_\l\log\det(-D_R^2)&\equiv&\int_0^\l\partial_\l\log\det(-D_R^2)\non\\
&=&\int_0^\l\partial_\l\Tr\log(-D_R^2)
\la{trlog}
\ea
There is a standard expression~\cite{Brown:1977eb} for the inverse $\tilde\Delta_R(x,y)$ of the Dalembertian:
\be
-D_R^2\tilde\Delta_R(x,y)=\d(x-y).
\la{prop}
\ee
This standard expression is given below for the HS caloron. But it has not the right (anti)-periodicity. 
This can be enforced by considering :
\be
\Delta_R^\eta(x,y)\equiv\sum_{n=-\infty}^{\infty}(\pm 1)^n\tilde\Delta_R(x,y+n/T),
\la{perprop}
\ee
 and this (anti)-periodic propagator still obeys Eq. (\ref{prop}) with an (anti-)periodic delta function.  The suffix $\eta$ refers to (anti)-periodicity.
 
 The variation in Eq. (\ref{trlog}) obeys then:
 \be
 \partial_\l\Tr_\eta \log(-D_R^2)=\sum_{n=-\infty}^{\infty}(\pm)^n\Tr(-\partial_\l D_R^2\tilde \Delta(x,y)_{y=x+n/T})=\mathrm{Tr}(-\partial_\l D_R^2\Delta^\eta(x,y))
\la{variationsumprop}
 \ee  
 
 The trace is integration over all space. In time we integrate only over one period. Furthermore a trace
 over color $\mathrm{Tr}_c$ is involved:
 \be
 \Tr\equiv \int_0^{1/T} dt\int  d\vec x ~\Tr_c=\int_{R^3\times S}  dx^4~\Tr_c.
 \la{deftrace}
 \ee

 The fundamental building block has a simple relation with the quantity $\hat Z_1$, 
 which is called the total instanton density. Before discussing this we have to turn to the thermal instanton in more detail.

\subsection{HS caloron properties}\la{sec:caloronprop}

In this section some of the  salient properties of the thermal instanton will be discussed. The periodic instanton
was introduced~\cite{Harrington:1978ve} on the basis of the multi-instanton solution of 't Hooft~\cite{'tHooft:unpublished}.
In our quaternion notation $x=x_\m\s_\m$, $x_m=x-m/T\s_0$ and  $x_m^2=x_mx_m^\dagger$,  it reads:     
\ba
 A^a_{\mu} &=&  -\bar\eta^a_{\mu\nu}\frac{\partial_\nu\Pi}{\Pi} \label{eq:gf} \\
 \Pi(r,t)&=&1+\rho^2\sum_m{x_mx_m^\dagger\over{x^2_mx^2_m}},~\mbox{or}\label{eq:pi11}\\
 \Pi(r,t)&=&1+{\pi\r^2T\over{r}}{\sinh(2\pi rT)\over {(\cosh(2\pi rT)-\cos(2\pi tT))}}\la{pi1}
\ea  
\noindent where $\bar\eta^a_{\m\n}$ follows from $\s_\m\s_\n^\dagger=\d_{\m\n}+i\t^a\bar\eta^a_{\m\n}$, and $\s_\m=({\bf 1},-i\vec \t)$.
In what follows we absorb the temperature $T$ into the space-time coordinates. The combination $\l\equiv\pi\r T$ measures the overlap 
between the instantons and therefore we write the prepotential $\Pi$ as:

\be
 \Pi(r,t)=1+{\l^2\over{\pi r}}{\sinh(2\pi r)\over {(\cosh(2\pi r)-\cos(2\pi t))}}\la{pi}
 \la{HSpot}
\ee
In what follows we will refer to this configuration as the HS-caloron.
In these units the single instanton (labeled by the suffix $0$) reads:
\be
 \Pi_0(r,t)=1+{\l^2\over{\pi^2(r^2+t^2)}}
 \la{pi0smalldsit} 
\ee
 At short distance $r,t \ll 1$    GPY noticed the HS potential  (\ref{HSpot}) behaves like:
\be
 \Pi(r,t)=1+{\l^2\over 3}+{\l^2\over{\pi^2(r^2+t^2)}}+\OO(r^2)
\ee
\noindent for any value of $\l$.
Hence the HS potential $A_\m^a$ behaves at short distance in term of the single instanton  function $\Pi_0$ like
\be
A^a_{\m}(x)=-\bar\eta^a_{\mu\nu}\partial_\nu\log\Pi_0(\l^2/(1+\l^2/3)).
\la{amushortdistance}
\ee
Note that the single instanton overlap $\l^2$ is reduced by a factor $1+{\l^2\over 3}$.
     
At large distance $r\gg\l$ the HS caloron behaves like a self-dual dipole:
\be
E^a_k=B^a_k\sim (\delta^a_k-3n^an_k)/r^3.
\ee 
The fields are rotational invariant if rotations and color transformations are in lockstep.
In the opposite limit, $\l\gg r\gg1$, the HS caloron looks like a self-dual monopole.

The HS instanton has eight periodic zero modes and how to deal with them is explained in GPY~\cite{Gross:1980br}. 
Their resulting expression for the total density with a Pauli-Villars scale $\Lambda$ is:
\be
{T\over V}\hat Z_1=(\pi T)^4\int d\l~n(\l,T/\Lambda).
\la{z1instantondensity}
\ee

This implies, using Eq. (\ref{totalfreeenergy}) and $f_{SB}=-p_{SB}$, that the correction to the Stefan-Boltzmann pressure is:
\be
p_{SB}+2(\pi T)^4\int d\l ~n(\l,T/\Lambda)
\la{correctionsb}
\ee

In this way the determinant contributes to the pressure.

In the remaining part of the paper we will compute $\log\det(-D_R^2)$.

\subsection{Propagators of isospin 1/2 and 1 in the field of the thermal instanton}\la{sec:propagators}

We need explicit propagators in the background field, according to Eq. (\ref{variationsumprop}). They have been given for isospin 1/2 and 1.
For isospin 1/2~\cite{Brown:1977eb}:  
\be
\tilde\Delta_{1/2}(x,y)=\Pi^{-1/2}(x){F_{1/2}(x,y)\over{4\pi^2(x-y)^2}}\Pi^{-1/2}(y)
\label{eq:prop12}
\ee
In the numerator we have the quaternion:

\begin{eqnarray}
 F_{1/2}(x,y) = 1 + \rho^2\sum_m \frac{x_m^\mu y^{\nu}_m  }{x_m^2 y^2_m}\sigma_\mu\bar\sigma_\nu
 \label{f12qq}
\end{eqnarray}
where $x_m = x - (m/T)\sigma_0$. 
Clearly this quaternion reduces to $\Pi$, Eq. (\ref{eq:pi11}), when the arguments coincide. 

$F_{1/2}(x,y)$  can also be written as:
\begin{eqnarray}
 F_{1/2}(x,y) = \FF_0(x,y) + i\tau_a \FF_a(x,y) \la{f120a}
 \end{eqnarray} 
where
 \begin{eqnarray}
 \FF_0(x,y) &=& 1 + \rho^2\sum_m \frac{x_m^\mu y^{\mu}_m  }{x_m^2 y^2_m} \\
 \FF_a(x,y) &=& \bar\eta_{a\mu\nu}\rho^2\sum_m \frac{x_m^\mu y^{\nu}_m  }{x_m^2 y^2_m} 
 \end{eqnarray}
For isospin 1~\cite{Brown:1977eb}:
\ba    
\tilde\Delta_1^{ab}(x,y)&=&\Pi^{-1}(x)\bigg[{F_1^{ab}(x,y)\over{4\pi^2(x-y)^2}}+{K_1^{ab}(x,y)\over{4\pi^2}}\Bigg]\Pi^{-1}(y)\non\\
F_1^{ab}(x,y)&=&{1\over 2}\mathrm{Tr}_c\left(\t^a F_{1/2}(x,y)\t^b F_{1/2}^\dagger(x,y)\la{f1}\right)
\ea
\noindent where $K_1^{ab}(x,y)$ is a non-singular matrix. 
All the singular behavior of the propagator is concentrated in the first term: as 
$x\rightarrow y$ the first term becomes just the free propagator  ${1\over{}4\pi^2(x-y)^2}$.  The form of the second term is entirely factorized:

\be
 K_1^{ab}(x,y)={\r^2\over 2}\Tr_c({x_m\over{x_m^2} }\t^a {x^\dagger_n\over{x_n^2}}f_{mn; r s} {\r^2\over 2}\Tr_c {y^\dagger_r\over{y_r^2}} \t^b {y_s\over{y_s^2}}.
\la{Csymmetric}
 \ee        
 
The two factors are both anti-symmetric in $m\leftrightarrow n$ and $r\leftrightarrow s$, and symmetric in simultaneous exchange $(m,n)\leftrightarrow (r,s)$. The
reason is the symmetry
\be
\Delta^{ab}(x,y)=\Delta^{ba}(y,x).
\ee

The matrix $f$ is in the HS case simply:
\be
f_{rs,tu}={\d_{rt}\d_{su}-\d_{ru}\d_{st}\over{(r-s)^2}}-\r^2{h_{rs,tu}\over{(r-s)^2(t-u)^2}}
\ee

The second term will give a vanishing contribution to the thermalized propagator, as was noted in GPY, Appendix E.

These propagators are the essential  input for the determinants.

\section{The Harrington-Shephard (HS) caloron contribution}\la{sec:HS}

In this section we give our result for the determinants.

The HS contribution was first computed in the seminal paper of Gross, Pisarski and Yaffe~\cite{Gross:1980br} (hereafter referred to as GPY ).
Their method was quite elegant for the isospin $1/2$ case. But this elegance was lost in the isospin 1 contribution, and their final result was
obtained by numerically integrating a huge number of terms in  the trace appearing in the logarithm of the determinant. Our result is equally simple for {\it both} cases.
We start by writing down the definitions: 
\begin{eqnarray}
\mathrm{Tr}\left( - \d_\l D^2_{1/2}\Delta^\eta_{1/2}\right) -\mathrm{Tr}\left( -\d_\l D^2_{0,1/2}\Delta_{0,1/2}\right)&=&   A(\lambda) + \eta\frac{\lambda^2}{3}  \\
\mathrm{Tr}\left( - \d_\l D^2_{1}\Delta^\eta_{1}\right)-\mathrm{Tr}\left( - \d_\l D^2_{0,1}\Delta_{0,1}\right) &=&  4  A(\lambda) + B_\eta(\lambda) +
\eta\frac{4}{3}\lambda^2.     
\la{determinants}
 \end{eqnarray}
 The shorthand $\d_\l$ is explained below Eq. (\ref{trlog}).
 
We introduce   $\eta = 1 (-1/2)$ for periodic (anti-periodic) boundary conditions multiplying the surface terms proportional to $\l^2$.  The anti-periodic case concerns zero mass Dirac fermions with a fixed helicity,  or equivalently zero mass Majorana fermions.
$\Delta^\eta_R$ is the thermal propagator obtained by (anti)-periodizing the propagator $\tilde\Delta_R$ constructed by Brown et al.~\cite{Brown:1977cm,Brown:1977eb,Brown:1978yj}. 
We subtracted the single instanton contribution (labelled by the suffix $0$)
to get rid of the logarithmic short distance singularity common to both contributions. This singularity is only present in the term where the propagators are 
taken at the same point and leads to the first equation below:
\begin{eqnarray}
 A(\lambda) &=&\frac{1}{12} \frac{1}{16\pi^2}\left( \int_{R^3\times S} dx^4\frac{((\partial_\mu \Pi)^2)^2}{\Pi^4} 
-\int_{R^4} dx^4\frac{((\partial_\mu \Pi_0)^2)^2}{\Pi_0^4}\right)   \nonumber \\
 B_\eta(\lambda) &=&  \frac{1}{16\pi^2} \int_{R^3\times S}  dx^4\frac{((\partial_\mu \Pi)^2)^2}{\Pi^4}H_\eta   
\la{determinantdetail}
\ea     
The second equation concerns all the terms we get by (anti)-periodizing the propagator. As we will see in section \ref{sec:determinants} it separates quite naturally into a surface term
proportional to $\l^2$ and a simple volume term proportional to $ B_\eta(\l)$~\footnote{
Note that  dependence on renormalization scheme does only come in through the calculation of the single instanton determinant. $A(\l)$ is unambiguous}.

The expression for the function $H_\eta$ is {\it independent} of the overlap variable $\l$ and is expressed in terms of the scalar potential of the monopole
\be
h(r)=\coth(2\pi r)-{1\over{2\pi r}}
\la{monopolescalar} 
\ee
and in terms of 
\be
f(r,t)={1\over{\pi r}}{\sinh(2\pi r)\over {\cosh(2\pi r)-\cos(2\pi t)}}
\ee 
appearing in the HS potential:
\be
\Pi=1+\l^2f(r,t).
\ee
For periodic boundary conditions one gets:
\be
H_p(r,t)={h^2(r)-1\over{h^2(r)-1+f(r,t)}}\la{periodichp},
\la{periodich}
\ee
\noindent and for anti-periodic boundary conditions:
\be
H_a(r,t)=\Bigg({\coth(2\pi r)\over{\sinh(2\pi r)}}+\Bigg({1\over{2\pi r}}\Bigg)^2-{1\over{\pi r \sinh (2\pi r)}}\Bigg)/(h(r)^2-1+f(r,t))
\la{antiperiodich}
\ee

The strategy to get these simple expressions for the thermal part of the HS determinant is explained in section \ref{sec:determinants}.

At this point it is instructive to see how the logarithmic behaviour in $\l$ follows from examining the behavior of the integrands. 

For $A(\l)$ the logarithmic behavior  is dictated by  short distance behavior of the traces leading to $ A(\l)$ in Eq. (\ref{determinantdetail}). This is clear from the behavior
of the HS caloron near its center, as discussed in section \ref{sec:caloronprop}:
\be 
\Pi(r,t)=1+{\l^2\over 3}+{\l^2\over {r^2+t^2}}+\OO(r^2).
\ee
This leads to a logarithmic divergence in both of the two members of the first equation of Eq. (\ref{determinantdetail}), but it cancels out in the difference, leaving a term $-{1\over{12}}\log(1+{\l^2\over 3})$. 

In contrast, for $ B_p(\l)$ the large $\l$ logarithm stems
from its {\it long} distance behavior. To understand this the following identity is quite useful:

\begin{eqnarray}
 \frac{ (\partial_\mu \Pi)^2}{\Pi^2}= 4\pi^2\frac{f^2}{(\frac{1}{\lambda^2} + f)^2}(h^2-1+f).
\la{explicitlambda}
 \end{eqnarray}
It follows in a straightforward way by differentiation of the left hand side and using the definitions of $f$ and $h$ above. It is useful because the 
overlap dependence $\l$ is explicit.

Substituted in the definition of $ B_p$, Eq.(\ref{determinantdetail}) and use of Eq. (\ref{periodich}),  the result is:
\be
 B_p(\l;r,t)= 16\pi^4{f^4\over{({1\over{\l^2}}+f)^4}}(h^2-1+f)(h^2-1).
\ee   
Setting ${1\over{\l^2}}=0$ gives a logarithmic infrared  divergence in $\int d\vec x dt  B_p(r,t)$. 
So both $ A$ and $ B_p$ have logaritmic behaviour in $\l$, though for different reasons. 
And indeed in a wide range of values of $\l$ we find numerically, as in GPY:
\be
 B_p(\l)=12 A(\l).
\ee
We obtained the fits (see fig. (\ref{fig:numint})):
\begin{eqnarray}
  B_p(\l) &=& -\log\left(1 + \frac{\lambda^2}{3}\right) + 12\alpha(1 + \gamma \lambda^{-3/2})^{-8}\nonumber\\
 \alpha &=& 0.012972(2), \,\,\,\ \gamma = 0.8995 (6)   
\la{barbfit}
\end{eqnarray}

This matches very well with the numerical  GPY result~\cite{Gross:1980br}. 
It is useful to note that the logarithmic term dominates in both cases 
for all values of the overlap. For vanishing $\l$ they are equal and for large values of the overlap $\l$ there is equality, i.e.  only the logarithm. 

For $ B_a(\l)$ no infrared divergence appears, as follows immediately from comparing the behaviour of the numerators in (\ref{antiperiodich}) and (\ref{periodich}): the first goes like 
${1\over r}$ compared to the second. The fit to its behavior reads (see fig. (\ref{fig:numint})):
\begin{eqnarray}
 B_a(\lambda) &=& \frac{a\lambda^2}{b + \lambda^2} + \frac{\bar\alpha}{(1+\bar\gamma \lambda^{-3/2})^8} . 
 \la{apfit}
 \end{eqnarray}
where:
\begin{eqnarray}
a = 0.550(2),\,\,\, b = 3.25(2) \\ 
\bar\alpha = 0.107(2),\,\,\, \bar\gamma = 1.22(2)
\end{eqnarray}

It remains tantalizing to understand why the bosonic thermal contribution $B_p(\l)$ is numerically so near to its
short distance partner $A(\l)$. 
 In the conclusion we offer some explanation, but basically we have no other 
understanding than what brute force teaches us in the next section \ref{sec:determinants}
\footnote{Beautiful papers based on the ADHM formalism offer deep insights, but did not help in our 
particular problem\cite{Corrigan:1979di}\cite{Christ:1978jy}\cite{Osborn:1979bx}\cite{Jack:1979rn}\cite{Berg:1979ku}\cite{Berg:1980wf}.}.
 \begin{figure}   
  \centering 
  \includegraphics{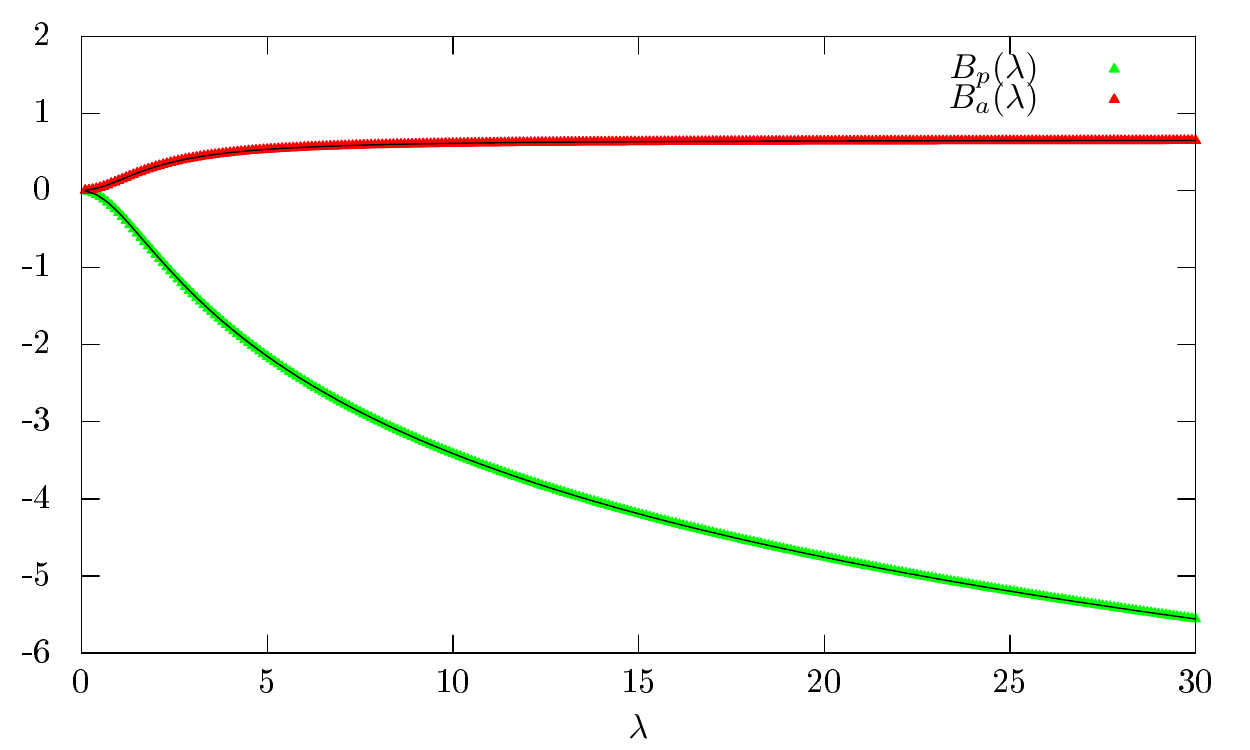}
  \caption{Periodic function $ B_p(\l)$ in green and antiperiodic  function $B_a(\l)$ in red. The black curves correspond to the fits in Eq. (\ref{barbfit}) and (\ref{apfit}). The plot of $A(\l)$  falls on top of that of $B_p(\l)$.  }
  \label{fig:numint}
 \end{figure}     
\subsection{Twisted gluinos}
In recent years, more general boundary 
conditions have been studied, i.e. \cite{Bilgici:2009jy,GarciaPerez:2009mg,Misumi:2014raa}. An example is given by fermions
with twisted boundary conditions:
\begin{eqnarray}
 \psi(\vec{x},t+n) = e^{i2\pi\alpha n}\psi(\vec{x},t)\,.
 \la{twistedgluino}
\end{eqnarray}
These boundary conditions can be seen as a smooth interpolation between the periodic (bosonic gauge particles) and anti-periodic case (Majorana fermions), because the degrees of freedom do not change. Only the thermodynamics of the latter two species is well defined.    
The periodization of the propagator is now given by:
\begin{eqnarray}
 \Delta_R^{\alpha}(x,y) = \sum_{n} e^{i2\pi \alpha n}\tilde\Delta_R(x,y-n)\,. 
\end{eqnarray}
Using our procedure, the generalization to this case is trivial.  Substituting Eq. (\ref{eq:sumn}) through (\ref{eq:sumcos}) in Eqs. (\ref{eq:sump})
and (\ref{eq:b2p}) we obtain:
\begin{eqnarray}
\mathrm{Tr}\left( - \d_\l D^2_{1/2}\Delta^\alpha_{1/2}\right) -\mathrm{Tr}\left( -\d_\l D^2_{0,1/2}\Delta_{0,1/2}\right)&=&   A(\lambda) +2\mathcal B_2(\alpha)\lambda^2  \\
\mathrm{Tr}\left( - \d_\l D^2_{1}\Delta^\alpha_{1}\right)-\mathrm{Tr}\left( - \d_\l D^2_{0,1}\Delta_{0,1}\right) &=&  4  A(\lambda) + B_\alpha(\lambda) +
8\mathcal B_2(\alpha)\lambda^2.     
\la{determinants}
 \end{eqnarray}
where
\begin{eqnarray}
\mathcal B_2(\alpha) &=&  \alpha^2 - \alpha + \frac{1}{6} \\
  B_\alpha(\lambda) &=&  \frac{1}{16\pi^2} \int_{R^3\times S}  dx^4\frac{((\partial_\mu \Pi)^2)^2}{\Pi^4}H_\alpha   \\
  H_\alpha &=& \frac{P_\alpha}{h^2 -1 + f} \\
  P_\alpha &=& -\frac{1}{2\pi r^2}\partial_r \left(r^2 h_\alpha(r)\right) \\
  h_\alpha &=& \frac{\cosh(2\pi r(1 - 2\alpha))}{\sinh(2\pi r)} - \frac{1}{2\pi r}
\end{eqnarray}
In Fig. \ref{fig:tbc} $B_\alpha(\lambda)$ is plotted for several values of $\alpha$.  The screening terms are given by the second order  Bernoulli polynomial $\mathcal B_2(\alpha)$, positive for $\alpha=0$, but negative for $\alpha=1/2$.
The negative screening mass
becomes positive when we consider the free energy for the fermionic case. For the intermediate cases we do not know how to assign a free energy. 

 \begin{figure}   
  \centering 
  \includegraphics{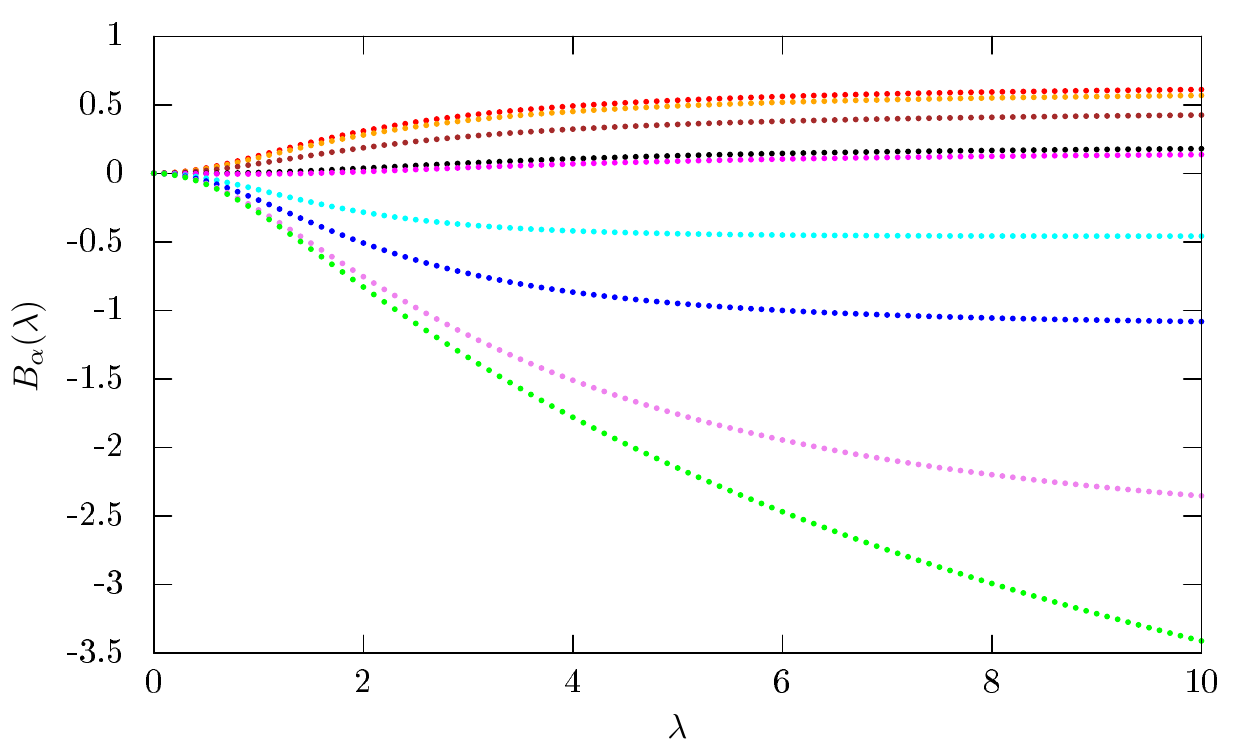}
  \caption{Thermal contribution from the twisted gluino, Eq. (\ref{twistedgluino}) with $\alpha = 0,0.01,0.05,0.1,0.2,1/2-\sqrt{3}/6,0.3,0.4,1/2$}
  \label{fig:tbc}
 \end{figure}

 \section{Determinants}
\label{sec:determinants}  
In this section, we start by explaining our strategy 
to compute the thermal contribution to the determinants for both 1/2 and 1 representations. 
Then, in the subsections the steps are made explicit.
 
Our main goal is computing the following quantity:
\begin{eqnarray}
 \mathrm{Tr}_n\left[ - \d_\l D^2_{R}\tilde\Delta_{R}\right] \label{eq:trnp}  
\end{eqnarray}
where $R = 1/2, 1$ denotes the representation and the operation $\mathrm{Tr}_n$ is defined by:
\begin{eqnarray}
  \mathrm{Tr}_n (X(x,y)) = \int dx^4\mathrm{Tr}_c X(x,x_n) 
\end{eqnarray}
where $\mathrm{Tr}_c$ denotes the trace in $su(2)$ matrix indices.  
Once we have this quantity, all we have to do is to sum over $n$ with the appropriate phase factor as in Eq. (\ref{twistedgluino}), with the help of \ref{app:sums}.

To get Eq.(\ref{eq:trnp}) (for $n$  fixed)  explicitly requires the handling of the sums over instantons, that constitute the caloron. This  leads to Eq.(\ref{firstresult}).  The isospin $1/2$ result follows then  immediately. The isospin one result is  contained in  (\ref{eq:f0fa}).

We start with the HS potential:
\begin{eqnarray}
 A^a_{\mu} &=&  -\bar\eta^a_{\mu\nu}\frac{\partial_\nu\Pi}{\Pi} \label{eq:gf} \\ 
\partial_\l A^a_{\mu} &=& -c_\l\bar\eta^a_{\mu\nu}\frac{\partial_\nu\Pi}{\Pi^2} \label{eq:vgf}\,,
\end{eqnarray}
where $\Pi$ is defined in Eq. (\ref{pi}) and $c_\l={2\over \l}$.    

The SU(2) propagator can be written in both representations as (see Eqs. (\ref{eq:prop12}) and (\ref{f1})):
\begin{eqnarray}    
 \tilde\Delta_R(x,y) = \frac{1}{4\pi^2}\Pi^{-R}(x)\left(\frac{F_R(x,y)}{(x-y)^2} + K_R(x,y)\right)\Pi^{-R}(y)\,.
\la{propagators}
\end{eqnarray}
Explicit expressions for $F_R$ and $K_R$ for both 1/2 and 1 representation are given in the next subsections.

The case $n=0$ was computed by Brown~\cite{Brown:1978yj} and GPY and the result is summarized in the last subsection. 
 Let us start with the derivative of the logarithm of the determinant:
\ba
\partial_\l \log\det (-D^2_R))&=&\Tr~\partial_\l(-D^2_R\tilde\Delta_R(x,y)_{y=x_n})\non\\
&=&\Tr\Bigg(2i\partial_\l A_{\m R}(x)\partial^x_\m\tilde\Delta(x,y)_{y=x_n}+\Bigg\{\partial_\l A_{\m R}(x),A_\m(x)\Bigg\}\tilde\Delta(x,x_n)   \Bigg)
\la{derivativelogdet}
\ea 
The total derivative:
\be
\Tr ~\partial^x_\m\Bigg(\partial_\l A_{\m R}(x)\tilde\Delta(x,x_n)\Bigg)=\Tr\partial_\l A_{\m R}(x)\Bigg(\partial^x_\m\tilde\Delta(x,y)_{y=x_n}+\partial^y_\m\tilde\Delta(x,y)_{y=x_n}\Bigg),
\ee
\noindent using the divergencelessness of $A_{\m R}$.
Substituting this into Eq. (\ref{derivativelogdet}) gives:
\ba
\partial_\l \log\det (-D^2_R))&=&\Tr~ i\partial^x_\m\Bigg(\partial_\l A_{\m R}(x)\tilde\Delta(x,x_n)\Bigg)\non\\&+&\Tr~i\Bigg(\partial_\l A_{\m R}(x)\partial^x_\m\tilde\Delta(x,y)_{y=x_n}-(\partial_\l A_{\m R}(x)\partial^y_\m\tilde\Delta(x,y)_{y=x_n}\Bigg)\non\\&+&\Bigg\{\partial_\l A_{\m R}(x),A_{\m R}(x)\Bigg\}\tilde\Delta_R(x,x_n)   \Bigg).
\ea
The first term is a total derivative, but only the color singlet in $\tilde\Delta$ has the correct large $|\vec x|$ behavior (the others fall off too fast), so the color trace vanishes. 
After working out the  color traces in the remaining terms and using Eq. (\ref{eq:gf}),  (\ref{eq:vgf}) and (\ref{propagators})
 we get the important result :
\begin{eqnarray}
 \mathrm{Tr}_n\left( - \partial_\l D^2_{R}\tilde\Delta_{R}\right) =\frac{c_\lambda}{4\pi^2}\int_{R^3\times S}  dx^4\left( 
  \frac{\partial_\nu \Pi}{\Pi^{2R+2}}  j_{\nu,R}(x,x_n) + 2\Omega_R \frac{(\partial_\nu \Pi)^2}{\Pi^{2R+3}}\mathrm{Tr}_{c}\left(\frac{F_R(x,x_n)}{n^2}
  + K_R(x,x_n)\right)\right)
\label{eq:det}
  \end{eqnarray}
where
\begin{eqnarray}
 j_{\nu,R}(x,x_n) &=& \bar\eta^a_{\mu\nu}\mathrm{Tr}_{c}\left(iT^a_R(\partial_\mu^x-\partial_\mu^y)
\left. \left(\frac{F_R(x,y)}{(x-y)^2} + K_R(x,y)\right)\right|_{y=x_n}\right) 
\label{eq:jmu}
 \end{eqnarray}
 and $\Omega_R =  T^a_RT^a_R$ is the quadratic Casimir operator.
 In the next subsection we compute separately the contribution from  $F_R(x,y)$ and  $K_R(x,y)$ and finally we combine
 both results and the contribution from $n=0$ to give the final result in the last subsection. 

 \subsection{Contribution from $F_R(x,y)$} 
Using  Eqs. (\ref{eq:f12q}) and (\ref{eq:f1}), $F_R(x,y)$ can be expanded as
\begin{eqnarray}
 F_{R}(x,y) =  F_{R0}(x,y){\bf 1}_R +  F_{Ra}(x,y)(iT_{Ra}) + F_{RS}(x,y),
\la{defF}
\end{eqnarray}
According to Eq. (\ref{eq:det}) we need:
\begin{eqnarray}
\mathrm{Tr}_{c}\left(\frac{F_R(x,x_n)}{n^2}\right) = \frac{1}{n^2}\left(
 d_RF_{R0}  + \mathrm{Tr}_cF_{RS}(x,x_n)\right) 
 \label{eq:tf}
\end{eqnarray}
where $d_R$ is the dimension of the representation. 

Now we turn to the term Eq. (\ref{eq:jmu}).
As we are interested in the sum over $n$, only the even part contributes.  
So using the results from Appendix \ref{app:fdf}, in particular Eq.(\ref{eq:df012}), (\ref{eq:df01}),  (\ref{eq:dfa12}) and (\ref{eq:dfa1})we obtain for the even part:
\begin{eqnarray}
 \bar\eta^a_{\mu\nu}\mathrm{Tr}_{c}\left((iT^a_{R})(iT^b_{R})(\partial_\mu^x-\partial_\mu^y)
\left. \left(\frac{F_{Rb}(x,y)}{(x-y)^2} \right)\right|_{y=x_n}\right) 
 = -\frac{2R}{n^2}\partial_\nu F_{R0}(x,x_n).
 \label{eq:jf}
\end{eqnarray}  

This is an important simplification, because on the right hand side only the function $F_{R0}$ defined in Eq.(\ref{defF}) appears. 
Inserting Eqs. (\ref{eq:tf}) and   (\ref{eq:jf}) in  (\ref{eq:det}) we obtain:
\begin{eqnarray} 
&& \mathrm{Tr}_n\left( - \partial_\l D^2_{R}\Delta_{R}\right)=\non\\
&& \frac{c_\lambda}{2\pi^2n^2}\int_{R^3\times S}  dx^4\left( 
  -\frac{\partial_\nu \Pi}{\Pi^{2R+2}} R\partial_\nu F_{R0}(x,x_n)
  + \Omega \frac{(\partial_\nu \Pi)^2}{\Pi^{2R+3}}(d_RF_{R0}(x,x_n)+ \mathrm{Tr}_c(F_{RS}(x,x_n))\right)
  \end{eqnarray} 
Integrating by parts the first term on the right hand side we  obtain:
\begin{eqnarray}
&& \frac{c_\lambda}{2\pi^2n^2}\frac{R}{2R+1}\int_{R^3\times S}  dx^4
  \partial_\nu \left(\partial_\nu\frac{1}{\Pi^{2R+1}} F_{R0}(x,x_n)\right) \nonumber \\
&+&c_\lambda\frac{d_R\Omega_R - R(2R+2) }{2\pi^2n^2}\int_{R^3\times S}  dx^4 
  \frac{(\partial_\nu \Pi)^2}{\Pi^{2R+3}} F_{R0}(x,x_n) \nonumber \\ &+&  
\frac{c_\lambda}{2\pi^2n^2}\int_{R^3\times S}  dx^4 
   \Omega_R \frac{(\partial_\nu \Pi)^2}{\Pi^{2R+3}} \mathrm{Tr}_c(F_{SR}(x,x_n))
 \la{firstresult}
  \end{eqnarray}
The first term is just a surface integral:
\begin{eqnarray}
c_\l \frac{1}{2\pi^2n^2}\frac{R}{2R+1}\int_{R^3\times S}  dx^4
  \partial_\nu \left(\partial_\nu\frac{ 1}{\Pi^{2R+1}} F_{R0}(x,x_n)\right) = 
 c_\l \frac{2R}{\pi^2n^2}\lambda^2.
\end{eqnarray}
The sums over $n$ are trivial and we obtain:
\begin{eqnarray}
\sum_{n\not=0}      c_\lambda \frac{2R}{\pi^2n^2}\lambda^2 &=& c_\lambda 2R\frac{\lambda^2}{3}\,, \label{eq:surp}\\
\sum_{n\not=0}(-1)^nc_\lambda \frac{2R}{\pi^2n^2}\lambda^2 &=& -c_\lambda 2R\frac{\lambda^2}{6}\,. \label{eq:sura}
\end{eqnarray}
Using Table \ref{tab:prop} it is easy to check that these surface terms are the only contribution for $R=1/2$, as was found
by GPY for the variation with respect to $\lambda$.  For example the volume term, the second one in Eq. (\ref{firstresult}), has a combination of group coefficients in front that
vanishes for $R=1/2$. 
\begin{table}
\centering
 \caption{Explicit formulas for  for $F_R$ in both spin 1/2 and 1 representations.
 $\FF_0$, $\FF_a$ are defined in 
 Eq. (\ref{f120a}),  $\vec\FF$ is a 3-vector which components are $\FF_a$ and $t$ denotes transposition to a column vector. }
\label{tab:prop}
\begin{tabular}{c|c|c|c|c|c}
 R & $d_R$ & $\Omega$ & $F_{R0}$ & $F_{RA}$ & $F_{RS}$ \\ \hline 
 1/2 & 2 & $\frac{3}{4}$ &$\FF_0$ & $2\FF_a$ &  0 \\
 1   & 3 & $2$ &$\FF_0^2-\FF_a^2$ & $2\FF_0 \FF_a$ & $2\vec \FF \vec \FF^t$
\end{tabular}      
\end{table}

The case $R=1$, can easily be obtained using the explicit forms of $F_{10}$ and $F_{1S}$ in Table \ref{tab:prop} :
\begin{eqnarray}
\Tr_n(-\partial_\l D^2_1)\Delta_1=\frac{c_\lambda }{\pi^2}\int dx^4 
  \frac{(\partial_\nu \Pi)^2}{\Pi^{5}}\frac{\FF_{0}^2(x,x_n)+ \FF_{a}^2(x,x_n)}{n^2} 
\la{eq:f0fa}  
\end{eqnarray} 
Finally, we perform the sums using Appendix \ref{app:sums}: 
 \begin{eqnarray} 
 \sum_{n\not=0} \frac{\FF_0^2(x,x_n) + \FF_a^2(x,x_n)}{n^2} &=& \frac{\pi^2}{3}\Pi^2 - 
 \frac{\pi}{2r}h(r)(\Pi^2-1)\,, \label{eq:sump}\\    
 \sum_{n\not=0}(-1)^n \frac{\FF_0^2(x,x_n) + \FF_a^2(x,x_n)}{n^2} &=& -\frac{\pi^2}{6}\Pi^2 - 
 \frac{\pi}{2r}h_a(r)(\Pi^2-1)\,, \label{eq:suma}
\end{eqnarray}
where
\begin{eqnarray}
 h(r) &=& \coth(2\pi r)-\frac{1}{2\pi r}\,,   \\
 h_a(r) &=& \frac{1}{\sinh(2\pi r)}-\frac{1}{2\pi r} \,.  
\la{monopolecharacter}
\end{eqnarray}  
The reader will recognize the  characteristic functions of the BPS monopole for respectively the scalar and vector potentials in Eq. (\ref{monopolecharacter}).

The contribution from the first term in Eqs. (\ref{eq:sump}) and (\ref{eq:suma}) are again surface integrals:
 \begin{eqnarray}
\eta\frac{c_\lambda}{3}\int_{R^3\times S}  dx^4  \frac{(\partial_\mu\Pi)^2}{\Pi^3}= c_\lambda\eta \frac{2 \lambda^2}{3 } \label{eq:sur2}
\end{eqnarray}
where $\eta=1(-1/2)$ for periodic (anti-periodic) boundary conditions. The remaining contribution 
is written for convenience in the following form:
\begin{eqnarray}
\partial_\l B_{1p}(\lambda) &=&  -{c_\lambda}\int_{R^3\times S}  dx^4 \frac{(\partial_\mu \Pi)^2}{\Pi^3}\frac{(\Pi-1)}{\Pi}\frac{1}{2\pi r}\left(h(r) + \frac{1}{\Pi}h(r)\right) 
\label{eq:b1p}
\end{eqnarray}
\begin{eqnarray}
\partial_\l B_{1a}(\lambda) &=&  -{c_\lambda}\int_{R^3\times S}  dx^4 \frac{(\partial_\mu \Pi)^2}{\Pi^3}\frac{(\Pi-1)}{\Pi}\frac{1}{2\pi r}\left(h_a(r) + \frac{1}{\Pi}h_a(r)\right) 
\label{eq:b1a}
\end{eqnarray}

Note that the second equation is obtained from the first by replacing $h$ by $h_a$.
\subsection{Contribution from $K_R(x,y)$}
Since $K_{1/2} = 0$ (see Eq. (\ref{eq:prop12})), this term only contributes to $R=1$. It was discussed in Section \ref{sec:propagators}, in particular Eq. (\ref{Csymmetric}).
Its contribution to the determinant was computed by GPY in \cite{Gross:1980br} and their result can be  written as~\footnote{The relevant formulas are in their Appendix E, underneath Eq. E.12}:
\begin{eqnarray}
\sum_n-\partial_\l D^2\left( \frac{1}{4\pi^2}\Pi^{-1}(x) K_1(x,y)\Pi^{-1}(y)|_{y=x_n}\right) &=& - \frac{c_\l}{\pi} \left(\frac{\partial_\nu \Pi}{\Pi^2}\frac{(\Pi-1)^2}{\Pi^2}
\partial_\nu \frac{1}{2r} h  +
  \frac{(\partial_\nu \Pi)^2}{\Pi^3}  \frac{(\Pi-1)^2}{\Pi^2}r \partial_r \frac{1}{2r} h\right). \nonumber
\la{k1ab}
\end{eqnarray}
Integrating by parts the first term, 
we obtain the contribution to the determinant:
\begin{eqnarray}
\partial_\l B_{2p}(\lambda) &=& - {c_\l}\int dx^4\frac{(\partial_\nu \Pi)^2}{\Pi^3}\frac{\Pi-1}{\Pi} \left( 
\frac{h}{\pi r} + r \partial_r\frac{h}{2\pi r} 
-  \frac{1}{\Pi}\left(\frac{2h}{\pi r} + r \partial_r \frac{h}{2\pi r}\right)\right)  
\label{eq:b2p}
\end{eqnarray} 
and for antiperiodic boundary conditions is easy to obtain, using the relevant formulas in  GPY and our Appendix \ref{app:sums}:
\begin{eqnarray}
\partial_\l B_{2a}(\lambda) &=&
- {c_\l}\int dx^4\frac{(\partial_\nu \Pi)^2}{\Pi^3}\frac{\Pi-1}{\Pi} \left( 
\frac{h_a}{\pi r} + r \partial_r\frac{h_a}{2\pi r}   
-  \frac{1}{\Pi}\left(\frac{2h_a}{\pi r} + r \partial_r \frac{h_a}{2\pi r}\right)\right)
\label{eq:b2a}    
\end{eqnarray}
As in Eq. (\ref{eq:b1a})  the second contribution is obtained from the first one by replacing $h$ by $h_a$.  In Appendix B we give a natural interpolation between  the two.
  
\subsection{Final results}
Combining Eqs. (\ref{eq:b1p}), (\ref{eq:b1a}), (\ref{eq:b2p}) and (\ref{eq:b2a}) we obtain:
\begin{eqnarray}
\partial_\l B_{\eta}(\lambda) &=&\partial_\l B_{1\eta}(\lambda) + \partial_\l B_{2\eta}(\lambda) =  \frac{c_\l}{4\pi^2}\int_{R^3\times S}  dx^4\frac{((\partial_\nu \Pi)^2)^2}{\Pi^5}
H_\eta(r,t)
\la{eq:finaltotal}
\end{eqnarray}
where 
\begin{eqnarray}
H_\eta(r,t) &=& \frac{P_\eta}{h^2-1 + f(r)} \\
P_p &=&  \frac{1}{\sinh(2\pi r)^2} + \frac{1}{4\pi^2r^2}-\frac{\coth(2\pi r)}{\pi r} \\
P_a &=&   \frac{\coth(2\pi r)}{\sinh(2\pi r)} + \frac{1}{4\pi^2r^2}-\frac{1}{\sinh(2\pi r)\pi r}
\end{eqnarray}
and we have used the relation:
\begin{eqnarray}
 (\Pi-1)^2 = \frac{1}{4\pi^2}(\partial_\mu\Pi)^2\frac{1}{h^2-1 + f(r)}
\end{eqnarray}
The contribution for $n=0$ can be easily computed using Eq. (3.24) in \cite{Brown:1978yj} and our Eqs. (\ref{eq:gf}) and (\ref{eq:vgf}):
\begin{eqnarray}  
  \partial_\l A(\lambda) = \frac{c_\l}{12}\frac{1}{4\pi^2}\left(
  \int_{R^3\times S^1} dx^4 \frac{((\partial_\mu \Pi)^2)^2}{\Pi^5}-\int_{R^4} dx^4 \frac{((\partial_\mu \Pi_0)^2)^2}{\Pi_0^5}\right). 
\end{eqnarray}  
Including the surface terms from Eqs (\ref{eq:surp}), (\ref{eq:sura}) and (\ref{eq:sur2}) we obtain our final result:
\begin{eqnarray}  
\mathrm{Tr}(-\partial_\l D_{1/2} \Delta^{\eta}_{1/2})-\mathrm{Tr}\left( -\partial_\l D^2_{0,1/2}\Delta_{0,1/2}\right) &=& \partial_\l A(\lambda) + \eta c_\l\frac{1}{3}\lambda^2 \\
\mathrm{Tr}(-\partial_\l D_{1} \Delta^{\eta}_{1})-\mathrm{Tr}\left( -\partial_\l D^2_{0,1}\Delta_{0,1}\right) &=& 4\partial_\l A(\lambda) + \partial_\l B_{\eta}(\lambda) + \eta c_\l\frac{4}{3}\lambda^2 
\end{eqnarray}  

The result for the integrated $A$ follows easily. That for the  $B_\eta$ follows as easily from the  $\l$-independence of $h$ and $f$ and we recover Eq. (\ref{determinantdetail}).

As we emphasized in Section \ref{sec:HS}, the final result for $B_{\eta}(\l)$ 
contains a factor $H_{\eta}$, {\it independent} of $\l.$. 
This factorization {\it fails} for the Eq. (\ref{eq:b1p}) and (\ref{eq:b2p}) 
individually. Only the sum of the two, Eq. (\ref{eq:finaltotal}) has this property. 
The same applies to the anti-periodic boundary conditions. 
There is therefore a subtle interplay between the corresponding terms in the isospin $1$ propagator (\ref{propagators}), $F_1(x,y)$ and $K_1(x,y)$.

\section{Conclusions} \la{sec:conclusions}
 
In this paper a surprisingly simple expression is given for the thermal ($n\neq 0$) part of the  isospin 1 determinants, and the corresponding free energies
for periodic and anti-periodic case.
      ~The former case is in very good agreement with the numerical result of GPY.  

The amount of analytic detail we needed  to establish  this simple result is almost embarrassing. It is clear we are far from understanding the underlying physics.

Most striking is the numerical equality of the $n=0$ result $4A(\l)$ and the  remaining thermal sum $B_p$.
 Note that the infra-red behaviour of the latter is related to the ultra-violet behaviour of the former.   
Naively this may have to do with the broken conformal invariance of the underlying theory. One may however argue that the short distance behavior is four dimensional and the  large distance behavior three dimensional because of the periodicity in time.  The periodicity plays an important role, because the equality is absent  for the anti-periodic case, see the upper curve in FIg. (\ref{fig:tbc}). A simple and intuitive understanding is very desirable.

We have also established an interpolation through twisted gluinos between the two results, for both isospin $1$ and isospin ${1\over 2}$. The isospin $1/2$ interpolates the screening term through  the second order Bernoulli polynomial.  The isospin $1$ case involves also $B_\a$, and it interpolates smoothly.  Unfortunately we do not know how  to assign a free energy to the interpolation.

 An example where the anti-periodic isospin $1$ case plays a role  is ${\cal{N}}=1$ theory. Here the  (n=0) part will drop out, and only the difference
 of  periodic $B_p$  and  anti-periodic $B_a$ will survive, together with the sum of the screening terms:
 \begin{eqnarray}
  n(\lambda,T/\Lambda)=n(T=0) e^{-(B_p(\lambda) - B_a(\lambda)) - 2\lambda^2}
 \end{eqnarray}
 
 The zero temperature density is given in GPY.
 From Fig. (\ref{fig:numint}) we see that the resulting density is only slightly higher.
 Nevertheless it remains quite  interesting to see in detail how our  result applies to the vanishing of the domain walls between the vacuum states of $\cal{N}=1$ theory.
 
 These domain walls separate the degenerate ground states . The latter are obtained by  applying transfomations of the discrete unbroken subgroup of the anomalous $U(1)$ group associated to the R-current $\bar\psi\g_\m\g_5\psi$~\cite{Kogan:1997dt}. The domain walls have an exactly calculable energy even in the strong coupling regime
and electric flux tubes can end on them, a property they share with branes.
As we heat up the system, the first transition will be, where the center group  symmetry is restored.  Going to even higher temperature we meet with the critical temperature where the fermion condensate vanishes. Then also  the non-anomalous subgroup becomes an unbroken symmetry in the plasma. Our instanton determinant is of course only a good approximation at even much higher temperatures.
An analysis of how R-symmetry is gradually restored in this phase is outside the scope of this paper is will be subject of  a later paper.  
 
 It is not hard to extend our results to other gauge groups, especially $SU(3)$, using the work of Christ et al.~\cite{Christ:1978jy}. 
 
 Most importantly, can it be extended to the case of calorons with non-trivial Polyakov loop\cite{Lee:1998bb,Kraan:1998pm}? This would
 be of interest for the effective potential, and could shed light on how it behaves when calorons are taken into account \cite{Diakonov:2004jn}, 
 ~\footnote{C.P. Korthals Altes and A. Sastre, in preparation}.

 \section{ Acknowledgements}
 We thank Robert Pisarski and Jan Smit for quite useful discussions and the referee for making some very constructive points.
 
 Both authors thank the Centre de Physique Th\'eorique for its hospitality during the beginning of this work.
CPKA is indebted to NIKHEF for hospitality. AS  
is funded by the DFG grant SFB/TR55 and thanks the Wuppertal theory group for its hospitality.

\appendix

\section{Determination of $F_R(x,y)$ and its derivatives}\la{app:fxyderivatives}
\label{app:fdf}  
In this Appendix, we compute the two important functions introduced in section \ref{sec:determinants}:
\begin{eqnarray}
&& \mathrm{Tr}_c(F_R(x,x_n)) \label{eq:f1} \\ 
&&\bar\eta^a_{\mu\nu}\mathrm{Tr}_c\left(iT^a(\partial_\mu^x - \partial_\mu^y)\frac{F_R(x,y)}{(x-y)^2}\right)_{y=x_n} \label{eq:f2}
 \end{eqnarray}
The fundamental matrix introduced by Brown ~\cite{Brown:1977eb} and used by GPY is:
\begin{eqnarray}
 F_{1/2}(x,y) = 1 + \rho^2\sum_m \frac{x_m^\mu y^{\nu}_m  }{x_m^2 y^2_m}\sigma_\mu\bar\sigma_\nu
\end{eqnarray}
where $x_m = x - m\sigma_0$. $F_{1/2}(x,y)$ can be written as a quaternion:
\begin{eqnarray}
 F_{1/2}(x,y) = \FF_0(x,y) + i\tau_a \FF_a(x,y) 
 \label{eq:f12q}
 \end{eqnarray}
where
 \begin{eqnarray}
 \FF_0(x,y) &=& 1 + \rho^2\sum_m \frac{x_m^\mu y^{\mu}_m  }{x_m^2 y^2_m} \\
 \FF_a(x,y) &=& \bar\eta^a_{\mu\nu}\rho^2\sum_m \frac{x_m^\mu y^{\nu}_m  }{x_m^2 y^2_m} 
 \end{eqnarray}
The components of $F_{1}(x,y)$ are given in terms of $F_{1/2}(x,y)$ by: 
\begin{eqnarray}
 F_{1}^{ab}(x,y) = \frac{1}{2}\mathrm{Tr}(\tau^a F_{1/2}(x,y)\tau^bF_{1/2}^\dagger(x,y))  
\end{eqnarray}  
using Eq. (\ref{eq:f12q}) we obtain:
\begin{eqnarray}    
 F_{1}(x,y) = (\FF_0^2(x,y) - \FF_a^2(x,y))I_{3\times3} + 2iT^a_{1}\FF_0(x,y)\FF_a(x,y) + 2 \vec{\FF}(x,y)\vec{\FF}^t(x,y)
 \label{eq:f1}
\end{eqnarray}
so combining Eq. (\ref{eq:f12q}) and (\ref{eq:f1}):
\begin{eqnarray}
 F_R(x,y) = F_{DR}(x,y)I_R + iT_{R}^aF_{Ra}(x,y) + F_{RS}(x,y)\,. 
\end{eqnarray}
In order to compute Eqs. (\ref{eq:f1}) and (\ref{eq:f2}),  we need to evaluate the case $y = x_n$. So, for Eq. (\ref{eq:f1}) we obtain \footnote{The sums used in this section are summarized in Appendix \ref{app:sums}}:
 \begin{eqnarray}
 \FF_0(x,x_n) &=& 1 + \rho^2\sum_m \frac{x_m^\mu x^{\mu}_{m+n}  }{x_m^2 x^2_{m+n}} = 1 + \frac{4r^2}{n^2+4r^2}(\Pi-1)\,. \label{eq:fn0} \\
 \FF_a(x,x_n) &=& \bar\eta^a_{\mu\nu}\rho^2\sum_m \frac{x_m^\mu x^{\nu}_{m+n}  }{x_m^2 x^2_{m+n}}= \rho^2\sum_m \frac{n x^a}{x_m^2x_{m+n}^2}=
\bar\eta^a_{0i} \frac{2nx^i}{n^2+4r^2}(\Pi-1)\,. \label{eq:nos}
 \end{eqnarray}
Eq. (\ref{eq:f2}) requires the evaluation of the difference of the derivatives in $y = x_n$. The determination of the  derivativates is a bit more laborious. 
We start with the derivative of $\FF_0(x,y)$:
\begin{eqnarray}
 \partial_\nu^x \FF_0(x,x_n)|_{y=x_n} &=& \rho^2\sum_m \frac{x_{m+n}^\nu}{x_n^2x_{n+m}^2}
 -2\rho^2\sum_m \frac{x_m^\mu x_{m+n}^\mu}{x_m^2x_{m+n}^2}\frac{x^\nu_m}{x_m^2}\,, \\
 \partial_\nu^y \FF_0(x,x_n)|_{y=x_n} &=& \rho^2\sum_m \frac{x_{m}^\nu}{x_n^2x_{n+m}^2}
 -2\rho^2\sum_m \frac{x_m^\mu x_{m+n}^\mu}{x_m^2x_{m+n}^2}\frac{x^\nu_{m+n}}{x_{m+n}^2}\,,  
\end{eqnarray}
so the difference is given by:
\begin{eqnarray}
 (\partial_\nu^x-\partial_\nu^y) \FF_0(x,x_n)|_{y=x_n} &=& \rho^2\sum_m \frac{x_{m+n}^\nu-x_{m}^\nu}{x_n^2x_{n+m}^2}
 -2\rho^2\sum_m \frac{x_m^\mu x_{m+n}^\mu}{x_m^2x_{m+n}^2}\left(\frac{x^\nu_m}{x_m^2} -\frac{x^\nu_{m+n}}{x_{m+n}^2}\right)\,. 
  \label{eq:dif0}
 \end{eqnarray}
In order to use the sums of Appendix \ref{app:sums} we rewrite the second term as:
\begin{eqnarray} 
 2\rho^2\sum_m \frac{x_m^\mu x_{m+n}^\mu}{x_m^2x_{m+n}^2}\left(\frac{x^\nu_m}{x_m^2} -\frac{x^\nu_{m+n}}{x_{m+n}^2}\right) =
  2\rho^2\sum_m \frac{x^\nu_m x_m^\mu x_{m+n}^\mu}{x_m^4}\frac{1}{x_{m+n}^2}
- 2\rho^2\sum_m \frac{x^\nu_{m+n}x_m^\mu x_{m+n}^\mu}{x_{m+n}^4}\frac{1}{x_m^2}  
\end{eqnarray}
and changing  $m+n \rightarrow m$ in the second sum we obtain:
\begin{eqnarray}
 2\rho^2\sum_m \frac{x_m^\mu x_{m+n}^\mu}{x_m^2x_{m+n}^2}\left(\frac{x^\nu_m}{x_m^2} -\frac{x^\nu_{m+n}}{x_{m+n}^2}\right) =
  2\rho^2\sum_m \frac{x^\nu_m }{x_m^4}\left(\frac{x_m^\mu x_{m+n}^\mu}{x_{m+n}^2}
- \frac{x_m^\mu x_{m-n}^\mu}{x_{m-n}^2}  \right).
\end{eqnarray}
Now, using the relations:
\begin{eqnarray}
 x_m^\mu x_{m+n}^\mu = x_m^2 - n(t-m),\ \\
 x_m^\mu x_{m-n}^\mu = x_m^2 + n(t-m),\
\end{eqnarray}
we obtain a more useful formula 
\begin{eqnarray}
 2\rho^2\sum_m \frac{x_m^\mu x_{m+n}^\mu}{x_m^2x_{m+n}^2}\left(\frac{x^\nu_m}{x_m^2} -\frac{x^\nu_{m+n}}{x_{m+n}^2}\right) =
   4n\left(\rho^2\sum_m \frac{x^\nu_m(t-m) }{x_m^2x_{m+n}^2x_{m-n}^2} 
-  n^2\rho^2\sum_m \frac{x^\nu_m(t-m) }{x_m^4x_{m+n}^2x_{m-n}^2}\right).
\label{eq:sect}
\end{eqnarray} 
We substitue Eq. (\ref{eq:sect}) in Eq. (\ref{eq:dif0}) and using sums in Appendix \ref{app:sums} as is explained in 
Eqs. (\ref{eq:adf012}) and (\ref{eq:adf01}) we obtain:
  \begin{eqnarray}
\bar\eta^a_{\nu\mu} (\partial_\nu^x-\partial_\nu^y) \FF_0(x,y)|_{y=x_n} &=&
- \bar\eta^a_{i\mu}  \partial_0 \FF_i(x,x_n) +
 \bar\eta_{a0\mu} \partial_r\left(\frac{2n r}{n^2+4r^2}(\Pi-1)\right)\label{eq:df012}  \\
\FF_a(x,x_n)\bar\eta_{a\nu\mu} (\partial_\nu^x-\partial_\nu^y) \FF_0(x,y)|_{y=x_n} &=&
 \frac{1}{2}\partial_\mu \FF_a^2(x,x_n)   \label{eq:df01}
\end{eqnarray}

The next step is computing the derivatives of 
 $\FF_a(x,x_n)$. As in the previous case we have:
\begin{eqnarray}
 \partial_\nu^x \FF_a(x,y)|_{y=x_n} &=&\phantom{-}\rho^2\sum_m \bar\eta^a_{\nu\alpha}\frac{ x_{m+n}^\alpha}{x_n^2x_{m+n}^2} - 2\rho^2\sum_m \bar\eta^a_{\alpha\beta}\frac{x_m^\alpha x_{m+n}^\beta}{x_n^2x_{n+m}^2}\frac{x_m^\nu}{x_m^2}\nonumber \\
 \partial_\nu^y \FF_a(x,y)|_{y=x_n} &=& -\rho^2\sum_m \bar\eta^a_{\nu\alpha}\frac{x_m^\alpha }{x_n^2x_{m+n}^2} - 2\rho^2\sum_m \bar\eta^a_{\alpha\beta}\frac{x_m^\alpha x_{m+n}^\beta}{x_n^2x_{n+m}^2}\frac{x_{m+n}^\nu}{x_{m+n}^2}\nonumber
\end{eqnarray}
and for  the difference
\begin{eqnarray}
 (\partial_\nu^x-\partial_\nu^y ) \FF_a(x,y)|_{y=x_n} &=&\rho^2\sum_m \bar\eta^a_{\nu\alpha}\frac{x_m^\alpha+ x_{m+n}^\alpha}{x_n^2x_{m+n}^2} - 2\rho^2\sum_m \bar\eta^a_{\alpha\beta}\frac{x_m^\alpha x_{m+n}^\beta}{x_n^2x_{n+m}^2}\left(\frac{x_m^\nu}{x_m^2} 
  - \frac{x_{m+n}^\nu}{x_{m+n}^2}\right). 
  \la{eq:sumfa}
\end{eqnarray}
Finally, multiplying by $\bar\eta$ and using Appendix \ref{app:sums} as is explained in Eqs. (\ref{eq:adfa12}) and (\ref{eq:adfa1}) we arrive at: 
\begin{eqnarray}
\bar\eta^a_{\nu\mu} (\partial_\nu^x-\partial_\nu^y ) \FF_a(x,y)|_{y=x_n} &=& - \partial_\mu \FF_0(x,x_n)+
\delta_{\mu a}\frac{4}{n}\FF_a(x,x_n) \label{eq:dfa12}   \\
\mathcal{F}_0(x,x_n)\bar\eta^a_{\nu\mu} (\partial_\nu^x-\partial_\nu^y ) \FF_a(x,y)|_{y=x_n} &=& 
- \frac{1}{2}\partial_\mu \FF_0^2(x,x_n)+
\delta_{\mu a}\frac{4}{n}\FF_0(x,x_n)\FF_a(x,x_n)  \label{eq:dfa1}   
\end{eqnarray}
Note that the second term in Eqs. (\ref{eq:dfa12}) and (\ref{eq:dfa1}) cancels with the derivative 
of $\frac{1}{(x-y)^2}$ in Eq. (\ref{eq:f2}).

\section{Sums}
\label{app:sums}
\subsection{Sums involving $m$, the location of instantons in time.}
We start remarking the relations:
\begin{eqnarray}
 x_m^\mu &=& (t-m,\vec x)\\
 x_m^\mu x^{\mu}_{m+n} &=& r^2 + (t-m)(t-m-n)
\end{eqnarray}
Using MATHEMATICA or another mathematical software is easy to check the following sums:
\begin{eqnarray}
 I_0(n) &=& \rho^2\sum_m\frac{1}{x_{m}^2x_{m+n}^2} = \frac{2}{n^2+4r^2}(\Pi -1 )\\
 I_1(n) &=& \rho^2\sum_m\frac{t-m}{x_{m}^2x_{m+n}^2} = \frac{n}{n^2+4r^2}(\Pi -1 )\\
 I_2(n) &=& \rho^2\sum_m\frac{(t-m)(t-m-n)}{x_{m}^2x_{m+n}^2} = \frac{2r^2}{n^2+4r^2}(\Pi -1 )\\
 I_3(n) &=& \rho^2\sum_m\frac{t-m}{x_{m}^2x_{m+n}^2x_{m-n}^2} = 0 \\
 I_4(n) &=& \rho^2\sum_m\frac{(t-m)^2}{x_{m}^2x_{m+n}^2x_{m-n}^2} = \frac{1}{2}\frac{1}{n^2+4r^2}(\Pi -1 )\\
 I_5(n) &=& \rho^2\sum_m\frac{t-m}{x_{m}^4x_{m+n}^2x_{m-n}^2} = -\frac{1}{2n^2}\partial_0\left(\frac{1}{n^2+4r^2}(\Pi-1)\right) \\
 I_6(n) &=& \rho^2\sum_m\frac{(t-m)^2}{x_{m}^4x_{m+n}^2x_{m-n}^2} = \frac{1}{2n^2}\frac{1}{n^2+4r^2}(\Pi -1 )
 \left(\frac{2n^2}{n^2+4r^2} + \frac{2\pi r (1-\cos(2\pi t)\cosh(2\pi r))}{\sinh(2\pi r)(\cosh(2\pi r) - \cos(2\pi t))}\right)\nonumber
 \end{eqnarray}
Using this sums we simplify Eqs. (\ref{eq:nos}), (\ref{eq:dif0}), (\ref{eq:sect}) and (\ref{eq:sumfa}) . For $F_{1/2}(x,x_n)$:
 \begin{eqnarray}
  \FF_0(x,x_n) &=& 1 + r^2 I_0(n) + I_3(n) = 1  + \frac{4r^2}{n^2+4r^2}(\Pi-1) \\
  \FF_a(x,x_n) &=& nx^a I_0(n)  = \frac{2n x^a}{n^2+4r^2}(\Pi-1)
 \end{eqnarray}
 For the difference of the derivatives for spin 1/2:
  \begin{eqnarray}
\bar\eta^a_{\nu\mu} (\partial_\nu^x-\partial_\nu^y) \FF_0(x,y)|_{y=x_n} &=& 
-\bar\eta^a_{0\mu}n\left(I_0(n)+4I_4(n) -  4 n^2I_6(m)\right)
 +4n^3\bar\eta^a_{i\mu}x^i   I_5(m) \nonumber \\
 &=&  \bar\eta^a_{0\mu} \partial_r\left(\frac{2n r}{n^2+4r^2}(\Pi-1)\right) - \bar\eta^a_{i\mu}
 \partial_0 \FF_i(x,x_n) 
\label{eq:adf012}\\
\bar\eta^a_{\nu\mu} (\partial_\nu^x-\partial_\nu^y ) \FF_a(x,y)|_{y=x_n} &=& 
\delta_{\mu a}4rI_0(n)  - \partial_\mu( r^2 I_0(n) + I_2(m)) \nonumber
\\  &=& -\partial_\mu \FF_0(x,x_n)  
+\delta_{\mu a} \frac{4}{n}\FF_a(x,x_n) \label{eq:adfa12}
\end{eqnarray}
For the difference of the derivatives for spin 1:
  \begin{eqnarray}
\FF_a(x,x_n)\bar\eta^a_{\nu\mu} (\partial_\nu^x-\partial_\nu^y) \FF_0(x,y)|_{y=x_n} &=&
 \frac{1}{2}\partial_\mu \FF_a^2(x,x_n) \label{eq:adf01} \\
\FF_0(x,x_n)\bar\eta^a_{\nu\mu} (\partial_\nu^x-\partial_\nu^y ) \FF_a(x,y)|_{y=x_n} &=& 
 -\frac{1}{2}\partial_\mu \FF_0^2(x,x_n)  
+\delta_{\mu a} \frac{4}{n}\FF_0(x,x_n)\FF_a(x,x_n) \label{eq:adfa1}
\end{eqnarray}

\subsection{Sums involving $n$,  the periodization of the propagators}
In order to (anti-)periodize the propagator, we have to substitute Eqs. (\ref{eq:fn0}) and (\ref{eq:nos}) in Eqs. (\ref{eq:sump}), (\ref{eq:suma}), 
(\ref{eq:b2p}) and (\ref{eq:b2a}). Then, the only dependence on $n$ appears as sums like:  
\begin{eqnarray}
\mathcal{I}_{a,b}^{(\eta)}(r) = \sum_{n\not = 0} (\pm1)^n\frac{1}{n^{2a}}\frac{1}{(n^2 + 4r^2)^b}.  
\end{eqnarray} 
where $\eta$ denotes periodic (+) and antiperidic (-) boundary conditions.
It is easy to realize that for $b > 0$:
\begin{eqnarray}
 \mathcal{I}_{a,b+1}^{(\eta)}(r) = -\frac{1}{8br}\partial_r  \mathcal{I}_{a,b}^{(\eta)}(r)\,.
\la{odd} 
 \end{eqnarray}
So we only need to compute  $\mathcal{I}_{a,1}^{(\eta)}$. 

We found a nice relation that works in all the cases we have
checked:
\begin{eqnarray}
\mathcal{I}_{a1}^{(\eta)}(r) = \sum_{n\not = 0} (\pm1)^n\frac{1}{n^{2a}}\frac{1}{n^2 + 4r^2} &=& \frac{(-1)^a}{(4r^2)^{a+1}}\mathrm{CoT}_{2a}\left(\hat h_{\eta}(r)\right)  
\label{eq:sumn}
\end{eqnarray}
where
\begin{eqnarray}  
\hat h_p(r) &=& 2\pi r \coth(2\pi r)\,, \\
\hat h_a(r) &=& \frac{2\pi r}{\sinh(2\pi r)}\,.
\end{eqnarray}
and we have define the  coTaylor operator at  order a, $\mathrm{CoT}_{a}$, by:
\begin{eqnarray}
\mathrm{CoT}_a(f(x)) \equiv f(x) - \mathrm{T}_a(f(x)) 
\end{eqnarray}
where  $\mathrm{T}_a(f(x))$ is just the Taylor expansion of $f(x)$ at order a. Unfortunatelly, we were not able to find
a formal proof of Eq. (\ref{eq:sumn}).  

The functions $h_\eta$ defined in Eqs. (\ref{monopolecharacter}) are related to $\hat h_{\pm}$ by:
\begin{eqnarray}
 h_\eta(r) = \frac{1}{2\pi r}\mathrm{CoT}_a(\hat h_\eta(r))\,.
\end{eqnarray}

Readers interested in more general boundary conditions, i.e.:            
\begin{eqnarray}
 \psi(\vec{x},t + 1) &=& e^{i2\pi\alpha}\psi(\vec{x},t)\,,   
\end{eqnarray}            
need to generalize Eqs. (\ref{eq:sump}) and (\ref{eq:suma}) to:
 \begin{eqnarray}
  \sum_{n\not=0}e^{i2\pi\alpha n} \frac{\FF_0^2(x,x_n) + \FF_a^2(x,x_n)}{n^2}\,.
 \end{eqnarray} 
and analogous for (\ref{eq:b2p}) and (\ref{eq:b2a}).

In  this case the only sums that appear are:
\begin{eqnarray}
\mathcal{I}_{a,b}^{(\alpha)}(r) = \sum_{n\not = 0} e^{i2\pi\alpha n}\frac{1}{n^{2a}}\frac{1}{(n^2 + 4r^2)^b}.  
\end{eqnarray} 
In this case, we found analogous formulas to Eqs. (\ref{eq:sumn}) (in all the cases we have checked):
\begin{eqnarray} 
\sum_{n\not = 0} \frac{1}{n^{2a}}\frac{\cos(2\pi \alpha n)}{n^2 + 4r^2} &=&
\frac{(-1)^a}{(4r^2)^{a+1}}\mathrm{CoT}_{2a}\left(\hat h_{\alpha}(r,\alpha)\right)\label{eq:sumcos}  \\
\sum_{n\not = 0} \frac{1}{n^{2a-1}}\frac{\sin(2\pi \alpha n)}{n^2 + 4r^2}
&=& \frac{(-1)^a}{(4r^2)^{a}}\mathrm{CoT}_{2a-1}\left(\hat s_{\alpha}(r,\alpha)\right)\label{eq:sumsin}  
\end{eqnarray}    
where 
\begin{eqnarray} 
\hat h_{\alpha}(r,\alpha) &=& 2\pi r \frac{\cosh(2\pi r(1 - 2\alpha))}{\sinh(2\pi r)}\,, \\ 
\hat s_{\alpha}(r,\alpha) &=& \pi  \frac{\sinh(2\pi r(1 - 2\alpha))}{\sinh(2\pi r)}\,,  
\end{eqnarray}    
for $0< \alpha < 1$. Eq. (\ref{eq:sumsin})   is only relevant voor non-trivial holonomy.  
Finally, in order to obtain the surface terms, the reader also needs the sum:
\begin{eqnarray}
\sum_{n\not = 0} \frac{\cos(2\pi \alpha n)}{n^{2}} &=&
 \pi^2\left(2\left(\frac{1}{2}-\alpha\right)^2 - \frac{1}{6}\right)\,.
\end{eqnarray}

\bibliography{caloron_prd}

\begin{thebibliography}{32}%
\makeatletter
\providecommand \@ifxundefined [1]{%
 \@ifx{#1\undefined}
}%
\providecommand \@ifnum [1]{%
 \ifnum #1\expandafter \@firstoftwo
 \else \expandafter \@secondoftwo
 \fi
}%
\providecommand \@ifx [1]{%
 \ifx #1\expandafter \@firstoftwo
 \else \expandafter \@secondoftwo
 \fi
}%
\providecommand \natexlab [1]{#1}%
\providecommand \enquote  [1]{``#1''}%
\providecommand \bibnamefont  [1]{#1}%
\providecommand \bibfnamefont [1]{#1}%
\providecommand \citenamefont [1]{#1}%
\providecommand \href@noop [0]{\@secondoftwo}%
\providecommand \href [0]{\begingroup \@sanitize@url \@href}%
\providecommand \@href[1]{\@@startlink{#1}\@@href}%
\providecommand \@@href[1]{\endgroup#1\@@endlink}%
\providecommand \@sanitize@url [0]{\catcode `\\12\catcode `\$12\catcode
  `\&12\catcode `\#12\catcode `\^12\catcode `\_12\catcode `\%12\relax}%
\providecommand \@@startlink[1]{}%
\providecommand \@@endlink[0]{}%
\providecommand \url  [0]{\begingroup\@sanitize@url \@url }%
\providecommand \@url [1]{\endgroup\@href {#1}{\urlprefix }}%
\providecommand \urlprefix  [0]{URL }%
\providecommand \Eprint [0]{\href }%
\providecommand \doibase [0]{http://dx.doi.org/}%
\providecommand \selectlanguage [0]{\@gobble}%
\providecommand \bibinfo  [0]{\@secondoftwo}%
\providecommand \bibfield  [0]{\@secondoftwo}%
\providecommand \translation [1]{[#1]}%
\providecommand \BibitemOpen [0]{}%
\providecommand \bibitemStop [0]{}%
\providecommand \bibitemNoStop [0]{.\EOS\space}%
\providecommand \EOS [0]{\spacefactor3000\relax}%
\providecommand \BibitemShut  [1]{\csname bibitem#1\endcsname}%
\let\auto@bib@innerbib\@empty
\bibitem [{\citenamefont {Tannenbaum}(2014)}]{Tannenbaum:2014tea}%
  \BibitemOpen
  \bibfield  {author} {\bibinfo {author} {\bibfnamefont {M.}~\bibnamefont
  {Tannenbaum}},\ }\href {\doibase 10.1142/S0217751X14300178} {\bibfield
  {journal} {\bibinfo  {journal} {Int.J.Mod.Phys.}\ }\textbf {\bibinfo {volume}
  {A29}},\ \bibinfo {pages} {1430017} (\bibinfo {year} {2014})},\ \Eprint
  {http://arxiv.org/abs/1406.1100} {arXiv:1406.1100 [nucl-ex]} \BibitemShut
  {NoStop}%
\bibitem [{\citenamefont {Szabo}(2014)}]{Szabo:2014iqa}%
  \BibitemOpen
  \bibfield  {author} {\bibinfo {author} {\bibfnamefont {K.}~\bibnamefont
  {Szabo}},\ }\href@noop {} {\bibfield  {journal} {\bibinfo  {journal} {PoS}\
  }\textbf {\bibinfo {volume} {LATTICE2013}},\ \bibinfo {pages} {014} (\bibinfo
  {year} {2014})},\ \Eprint {http://arxiv.org/abs/1401.4192} {arXiv:1401.4192
  [hep-lat]} \BibitemShut {NoStop}%
\bibitem [{\citenamefont {'t~Hooft}(1976{\natexlab{a}})}]{'tHooft:1976up}%
  \BibitemOpen
  \bibfield  {author} {\bibinfo {author} {\bibfnamefont {G.}~\bibnamefont
  {'t~Hooft}},\ }\href {\doibase 10.1103/PhysRevLett.37.8} {\bibfield
  {journal} {\bibinfo  {journal} {Phys.Rev.Lett.}\ }\textbf {\bibinfo {volume}
  {37}},\ \bibinfo {pages} {8} (\bibinfo {year}
  {1976}{\natexlab{a}})}\BibitemShut {NoStop}%
\bibitem [{\citenamefont {'t~Hooft}(1976{\natexlab{b}})}]{'tHooft:1976fv}%
  \BibitemOpen
  \bibfield  {author} {\bibinfo {author} {\bibfnamefont {G.}~\bibnamefont
  {'t~Hooft}},\ }\href {\doibase 10.1103/PhysRevD.18.2199.3,
  10.1103/PhysRevD.14.3432} {\bibfield  {journal} {\bibinfo  {journal}
  {Phys.Rev.}\ }\textbf {\bibinfo {volume} {D14}},\ \bibinfo {pages} {3432}
  (\bibinfo {year} {1976}{\natexlab{b}})}\BibitemShut {NoStop}%
\bibitem [{\citenamefont {Belavin}\ \emph {et~al.}(1975)\citenamefont
  {Belavin}, \citenamefont {Polyakov}, \citenamefont {Schwartz},\ and\
  \citenamefont {Tyupkin}}]{Belavin:1975fg}%
  \BibitemOpen
  \bibfield  {author} {\bibinfo {author} {\bibfnamefont {A.}~\bibnamefont
  {Belavin}}, \bibinfo {author} {\bibfnamefont {A.~M.}\ \bibnamefont
  {Polyakov}}, \bibinfo {author} {\bibfnamefont {A.}~\bibnamefont {Schwartz}},
  \ and\ \bibinfo {author} {\bibfnamefont {Y.}~\bibnamefont {Tyupkin}},\ }\href
  {\doibase 10.1016/0370-2693(75)90163-X} {\bibfield  {journal} {\bibinfo
  {journal} {Phys.Lett.}\ }\textbf {\bibinfo {volume} {B59}},\ \bibinfo {pages}
  {85} (\bibinfo {year} {1975})}\BibitemShut {NoStop}%
\bibitem [{\citenamefont {Shuryak}(1988)}]{Shuryak:1987jb}%
  \BibitemOpen
  \bibfield  {author} {\bibinfo {author} {\bibfnamefont {E.~V.}\ \bibnamefont
  {Shuryak}},\ }\href {\doibase 10.1016/0550-3213(88)90188-5} {\bibfield
  {journal} {\bibinfo  {journal} {Nucl.Phys.}\ }\textbf {\bibinfo {volume}
  {B302}},\ \bibinfo {pages} {559} (\bibinfo {year} {1988})}\BibitemShut
  {NoStop}%
\bibitem [{\citenamefont {Harrington}\ and\ \citenamefont
  {Shepard}(1978)}]{Harrington:1978ve}%
  \BibitemOpen
  \bibfield  {author} {\bibinfo {author} {\bibfnamefont {B.~J.}\ \bibnamefont
  {Harrington}}\ and\ \bibinfo {author} {\bibfnamefont {H.~K.}\ \bibnamefont
  {Shepard}},\ }\href {\doibase 10.1103/PhysRevD.17.2122} {\bibfield  {journal}
  {\bibinfo  {journal} {Phys.Rev.}\ }\textbf {\bibinfo {volume} {D17}},\
  \bibinfo {pages} {2122} (\bibinfo {year} {1978})}\BibitemShut {NoStop}%
\bibitem [{\citenamefont {Gross}\ \emph {et~al.}(1981)\citenamefont {Gross},
  \citenamefont {Pisarski},\ and\ \citenamefont {Yaffe}}]{Gross:1980br}%
  \BibitemOpen
  \bibfield  {author} {\bibinfo {author} {\bibfnamefont {D.~J.}\ \bibnamefont
  {Gross}}, \bibinfo {author} {\bibfnamefont {R.~D.}\ \bibnamefont {Pisarski}},
  \ and\ \bibinfo {author} {\bibfnamefont {L.~G.}\ \bibnamefont {Yaffe}},\
  }\href {\doibase 10.1103/RevModPhys.53.43} {\bibfield  {journal} {\bibinfo
  {journal} {Rev.Mod.Phys.}\ }\textbf {\bibinfo {volume} {53}},\ \bibinfo
  {pages} {43} (\bibinfo {year} {1981})}\BibitemShut {NoStop}%
\bibitem [{\citenamefont {Lee}\ and\ \citenamefont {Lu}(1998)}]{Lee:1998bb}%
  \BibitemOpen
  \bibfield  {author} {\bibinfo {author} {\bibfnamefont {K.-M.}\ \bibnamefont
  {Lee}}\ and\ \bibinfo {author} {\bibfnamefont {C.-h.}\ \bibnamefont {Lu}},\
  }\href {\doibase 10.1103/PhysRevD.58.025011} {\bibfield  {journal} {\bibinfo
  {journal} {Phys.Rev.}\ }\textbf {\bibinfo {volume} {D58}},\ \bibinfo {pages}
  {025011} (\bibinfo {year} {1998})},\ \Eprint
  {http://arxiv.org/abs/hep-th/9802108} {arXiv:hep-th/9802108 [hep-th]}
  \BibitemShut {NoStop}%
\bibitem [{\citenamefont {Kraan}\ and\ \citenamefont {van
  Baal}(1998)}]{Kraan:1998pm}%
  \BibitemOpen
  \bibfield  {author} {\bibinfo {author} {\bibfnamefont {T.~C.}\ \bibnamefont
  {Kraan}}\ and\ \bibinfo {author} {\bibfnamefont {P.}~\bibnamefont {van
  Baal}},\ }\href {\doibase 10.1016/S0550-3213(98)00590-2} {\bibfield
  {journal} {\bibinfo  {journal} {Nucl.Phys.}\ }\textbf {\bibinfo {volume}
  {B533}},\ \bibinfo {pages} {627} (\bibinfo {year} {1998})},\ \Eprint
  {http://arxiv.org/abs/hep-th/9805168} {arXiv:hep-th/9805168 [hep-th]}
  \BibitemShut {NoStop}%
\bibitem [{Note1()}]{Note1}%
  \BibitemOpen
  \bibinfo {note} {We use the word caloron for those periodic instantons which
  have trivial or non-trivial holonomy, i.e. they are spatially asymptoting
  into a (non-)trivial Polyakov loop. The case considered here with trivial
  Polyakov loop is called the thermal instanton, or HS caloron.}\BibitemShut
  {Stop}%
\bibitem [{\citenamefont {Diakonov}\ \emph {et~al.}(2004)\citenamefont
  {Diakonov}, \citenamefont {Gromov}, \citenamefont {Petrov},\ and\
  \citenamefont {Slizovskiy}}]{Diakonov:2004jn}%
  \BibitemOpen
  \bibfield  {author} {\bibinfo {author} {\bibfnamefont {D.}~\bibnamefont
  {Diakonov}}, \bibinfo {author} {\bibfnamefont {N.}~\bibnamefont {Gromov}},
  \bibinfo {author} {\bibfnamefont {V.}~\bibnamefont {Petrov}}, \ and\ \bibinfo
  {author} {\bibfnamefont {S.}~\bibnamefont {Slizovskiy}},\ }\href {\doibase
  10.1103/PhysRevD.70.036003} {\bibfield  {journal} {\bibinfo  {journal}
  {Phys.Rev.}\ }\textbf {\bibinfo {volume} {D70}},\ \bibinfo {pages} {036003}
  (\bibinfo {year} {2004})},\ \Eprint {http://arxiv.org/abs/hep-th/0404042}
  {arXiv:hep-th/0404042 [hep-th]} \BibitemShut {NoStop}%
\bibitem [{\citenamefont {Cherman}\ \emph {et~al.}(2014)\citenamefont
  {Cherman}, \citenamefont {Dorigoni},\ and\ \citenamefont
  {Unsal}}]{Cherman:2014ofa}%
  \BibitemOpen
  \bibfield  {author} {\bibinfo {author} {\bibfnamefont {A.}~\bibnamefont
  {Cherman}}, \bibinfo {author} {\bibfnamefont {D.}~\bibnamefont {Dorigoni}}, \
  and\ \bibinfo {author} {\bibfnamefont {M.}~\bibnamefont {Unsal}},\
  }\href@noop {} {\  (\bibinfo {year} {2014})},\ \Eprint
  {http://arxiv.org/abs/1403.1277} {arXiv:1403.1277 [hep-th]} \BibitemShut
  {NoStop}%
\bibitem [{\citenamefont {Brown}\ \emph {et~al.}(1977)\citenamefont {Brown},
  \citenamefont {Carlitz}, \citenamefont {Creamer},\ and\ \citenamefont
  {Lee}}]{Brown:1977cm}%
  \BibitemOpen
  \bibfield  {author} {\bibinfo {author} {\bibfnamefont {L.~S.}\ \bibnamefont
  {Brown}}, \bibinfo {author} {\bibfnamefont {R.~D.}\ \bibnamefont {Carlitz}},
  \bibinfo {author} {\bibfnamefont {D.~B.}\ \bibnamefont {Creamer}}, \ and\
  \bibinfo {author} {\bibfnamefont {C.-k.}\ \bibnamefont {Lee}},\ }\href
  {\doibase 10.1016/0370-2693(77)90515-9} {\bibfield  {journal} {\bibinfo
  {journal} {Phys.Lett.}\ }\textbf {\bibinfo {volume} {B70}},\ \bibinfo {pages}
  {180} (\bibinfo {year} {1977})}\BibitemShut {NoStop}%
\bibitem [{\citenamefont {Brown}\ \emph {et~al.}(1978)\citenamefont {Brown},
  \citenamefont {Carlitz}, \citenamefont {Creamer},\ and\ \citenamefont
  {Lee}}]{Brown:1977eb}%
  \BibitemOpen
  \bibfield  {author} {\bibinfo {author} {\bibfnamefont {L.~S.}\ \bibnamefont
  {Brown}}, \bibinfo {author} {\bibfnamefont {R.~D.}\ \bibnamefont {Carlitz}},
  \bibinfo {author} {\bibfnamefont {D.~B.}\ \bibnamefont {Creamer}}, \ and\
  \bibinfo {author} {\bibfnamefont {C.-k.}\ \bibnamefont {Lee}},\ }\href
  {\doibase 10.1103/PhysRevD.17.1583} {\bibfield  {journal} {\bibinfo
  {journal} {Phys.Rev.}\ }\textbf {\bibinfo {volume} {D17}},\ \bibinfo {pages}
  {1583} (\bibinfo {year} {1978})}\BibitemShut {NoStop}%
\bibitem [{\citenamefont {Brown}\ and\ \citenamefont
  {Creamer}(1978)}]{Brown:1978yj}%
  \BibitemOpen
  \bibfield  {author} {\bibinfo {author} {\bibfnamefont {L.~S.}\ \bibnamefont
  {Brown}}\ and\ \bibinfo {author} {\bibfnamefont {D.~B.}\ \bibnamefont
  {Creamer}},\ }\href {\doibase 10.1103/PhysRevD.18.3695} {\bibfield  {journal}
  {\bibinfo  {journal} {Phys.Rev.}\ }\textbf {\bibinfo {volume} {D18}},\
  \bibinfo {pages} {3695} (\bibinfo {year} {1978})}\BibitemShut {NoStop}%
\bibitem [{\citenamefont {'t~Hooft}()}]{'tHooft:unpublished}%
  \BibitemOpen
  \bibfield  {author} {\bibinfo {author} {\bibfnamefont {G.}~\bibnamefont
  {'t~Hooft}},\ }\href@noop {} {\bibinfo  {journal} {{unpublished}}\
  }\BibitemShut {NoStop}%
\bibitem [{Note2()}]{Note2}%
  \BibitemOpen
\bibfield  {journal} {  }\bibinfo {note} {Note that dependence on
  renormalization scheme does only come in through the calculation of the
  single instanton determinant. $A(\lambda )$ is unambiguous}\BibitemShut
  {NoStop}%
\bibitem [{Note3()}]{Note3}%
  \BibitemOpen
  \bibinfo {note} {Beautiful papers based on the ADHM formalism offer deep
  insights, but did not help in our particular problem\cite
  {Corrigan:1979di}\cite {Christ:1978jy}\cite {Osborn:1979bx}\cite
  {Jack:1979rn}\cite {Berg:1979ku}\cite {Berg:1980wf}.}\BibitemShut {Stop}%
\bibitem [{\citenamefont {Bilgici}\ \emph {et~al.}(2009)\citenamefont
  {Bilgici}, \citenamefont {Gattringer}, \citenamefont {Ilgenfritz},\ and\
  \citenamefont {Maas}}]{Bilgici:2009jy}%
  \BibitemOpen
  \bibfield  {author} {\bibinfo {author} {\bibfnamefont {E.}~\bibnamefont
  {Bilgici}}, \bibinfo {author} {\bibfnamefont {C.}~\bibnamefont {Gattringer}},
  \bibinfo {author} {\bibfnamefont {E.-M.}\ \bibnamefont {Ilgenfritz}}, \ and\
  \bibinfo {author} {\bibfnamefont {A.}~\bibnamefont {Maas}},\ }\href {\doibase
  10.1088/1126-6708/2009/11/035} {\bibfield  {journal} {\bibinfo  {journal}
  {JHEP}\ }\textbf {\bibinfo {volume} {0911}},\ \bibinfo {pages} {035}
  (\bibinfo {year} {2009})},\ \Eprint {http://arxiv.org/abs/0904.3450}
  {arXiv:0904.3450 [hep-lat]} \BibitemShut {NoStop}%
\bibitem [{\citenamefont {Garcia~Perez}\ \emph {et~al.}(2009)\citenamefont
  {Garcia~Perez}, \citenamefont {Gonzalez-Arroyo},\ and\ \citenamefont
  {Sastre}}]{GarciaPerez:2009mg}%
  \BibitemOpen
  \bibfield  {author} {\bibinfo {author} {\bibfnamefont {M.}~\bibnamefont
  {Garcia~Perez}}, \bibinfo {author} {\bibfnamefont {A.}~\bibnamefont
  {Gonzalez-Arroyo}}, \ and\ \bibinfo {author} {\bibfnamefont {A.}~\bibnamefont
  {Sastre}},\ }\href {\doibase 10.1088/1126-6708/2009/06/065} {\bibfield
  {journal} {\bibinfo  {journal} {JHEP}\ }\textbf {\bibinfo {volume} {0906}},\
  \bibinfo {pages} {065} (\bibinfo {year} {2009})},\ \Eprint
  {http://arxiv.org/abs/0905.0645} {arXiv:0905.0645 [hep-th]} \BibitemShut
  {NoStop}%
\bibitem [{\citenamefont {Misumi}\ and\ \citenamefont
  {Kanazawa}(2014)}]{Misumi:2014raa}%
  \BibitemOpen
  \bibfield  {author} {\bibinfo {author} {\bibfnamefont {T.}~\bibnamefont
  {Misumi}}\ and\ \bibinfo {author} {\bibfnamefont {T.}~\bibnamefont
  {Kanazawa}},\ }\href@noop {} {\  (\bibinfo {year} {2014})},\ \Eprint
  {http://arxiv.org/abs/1405.3113} {arXiv:1405.3113 [hep-ph]} \BibitemShut
  {NoStop}%
\bibitem [{Note4()}]{Note4}%
  \BibitemOpen
  \bibinfo {note} {The relevant formulas are in their Appendix E, underneath
  Eq. E.12}\BibitemShut {NoStop}%
\bibitem [{\citenamefont {Kogan}\ \emph {et~al.}(1998)\citenamefont {Kogan},
  \citenamefont {Kovner},\ and\ \citenamefont {Shifman}}]{Kogan:1997dt}%
  \BibitemOpen
  \bibfield  {author} {\bibinfo {author} {\bibfnamefont {I.~I.}\ \bibnamefont
  {Kogan}}, \bibinfo {author} {\bibfnamefont {A.}~\bibnamefont {Kovner}}, \
  and\ \bibinfo {author} {\bibfnamefont {M.~A.}\ \bibnamefont {Shifman}},\
  }\href {\doibase 10.1103/PhysRevD.57.5195} {\bibfield  {journal} {\bibinfo
  {journal} {Phys.Rev.}\ }\textbf {\bibinfo {volume} {D57}},\ \bibinfo {pages}
  {5195} (\bibinfo {year} {1998})},\ \Eprint
  {http://arxiv.org/abs/hep-th/9712046} {arXiv:hep-th/9712046 [hep-th]}
  \BibitemShut {NoStop}%
\bibitem [{\citenamefont {Christ}\ \emph {et~al.}(1978)\citenamefont {Christ},
  \citenamefont {Weinberg},\ and\ \citenamefont {Stanton}}]{Christ:1978jy}%
  \BibitemOpen
  \bibfield  {author} {\bibinfo {author} {\bibfnamefont {N.~H.}\ \bibnamefont
  {Christ}}, \bibinfo {author} {\bibfnamefont {E.~J.}\ \bibnamefont
  {Weinberg}}, \ and\ \bibinfo {author} {\bibfnamefont {N.~K.}\ \bibnamefont
  {Stanton}},\ }\href {\doibase 10.1103/PhysRevD.18.2013} {\bibfield  {journal}
  {\bibinfo  {journal} {Phys.Rev.}\ }\textbf {\bibinfo {volume} {D18}},\
  \bibinfo {pages} {2013} (\bibinfo {year} {1978})}\BibitemShut {NoStop}%
\bibitem [{Note5()}]{Note5}%
  \BibitemOpen
  \bibinfo {note} {C.P. Korthals Altes and A. Sastre, in
  preparation}\BibitemShut {NoStop}%
\bibitem [{Note6()}]{Note6}%
  \BibitemOpen
  \bibinfo {note} {The sums used in this section are summarized in Appendix
  \ref {app:sums}}\BibitemShut {NoStop}%
\bibitem [{\citenamefont {Corrigan}\ \emph {et~al.}(1979)\citenamefont
  {Corrigan}, \citenamefont {Goddard}, \citenamefont {Osborn},\ and\
  \citenamefont {Templeton}}]{Corrigan:1979di}%
  \BibitemOpen
  \bibfield  {author} {\bibinfo {author} {\bibfnamefont {E.}~\bibnamefont
  {Corrigan}}, \bibinfo {author} {\bibfnamefont {P.}~\bibnamefont {Goddard}},
  \bibinfo {author} {\bibfnamefont {H.}~\bibnamefont {Osborn}}, \ and\ \bibinfo
  {author} {\bibfnamefont {S.}~\bibnamefont {Templeton}},\ }\href {\doibase
  10.1016/0550-3213(79)90346-8} {\bibfield  {journal} {\bibinfo  {journal}
  {Nucl.Phys.}\ }\textbf {\bibinfo {volume} {B159}},\ \bibinfo {pages} {469}
  (\bibinfo {year} {1979})}\BibitemShut {NoStop}%
\bibitem [{\citenamefont {Osborn}(1979)}]{Osborn:1979bx}%
  \BibitemOpen
  \bibfield  {author} {\bibinfo {author} {\bibfnamefont {H.}~\bibnamefont
  {Osborn}},\ }\href {\doibase 10.1016/0550-3213(79)90347-X} {\bibfield
  {journal} {\bibinfo  {journal} {Nucl.Phys.}\ }\textbf {\bibinfo {volume}
  {B159}},\ \bibinfo {pages} {497} (\bibinfo {year} {1979})}\BibitemShut
  {NoStop}%
\bibitem [{\citenamefont {Jack}(1980)}]{Jack:1979rn}%
  \BibitemOpen
  \bibfield  {author} {\bibinfo {author} {\bibfnamefont {I.}~\bibnamefont
  {Jack}},\ }\href {\doibase 10.1016/0550-3213(80)90298-9} {\bibfield
  {journal} {\bibinfo  {journal} {Nucl.Phys.}\ }\textbf {\bibinfo {volume}
  {B174}},\ \bibinfo {pages} {526} (\bibinfo {year} {1980})}\BibitemShut
  {NoStop}%
\bibitem [{\citenamefont {Berg}\ and\ \citenamefont
  {Luscher}(1979)}]{Berg:1979ku}%
  \BibitemOpen
  \bibfield  {author} {\bibinfo {author} {\bibfnamefont {B.}~\bibnamefont
  {Berg}}\ and\ \bibinfo {author} {\bibfnamefont {M.}~\bibnamefont {Luscher}},\
  }\href {\doibase 10.1016/0550-3213(79)90061-0} {\bibfield  {journal}
  {\bibinfo  {journal} {Nucl.Phys.}\ }\textbf {\bibinfo {volume} {B160}},\
  \bibinfo {pages} {281} (\bibinfo {year} {1979})}\BibitemShut {NoStop}%
\bibitem [{\citenamefont {Berg}\ and\ \citenamefont
  {Stehr}(1980)}]{Berg:1980wf}%
  \BibitemOpen
  \bibfield  {author} {\bibinfo {author} {\bibfnamefont {B.}~\bibnamefont
  {Berg}}\ and\ \bibinfo {author} {\bibfnamefont {J.}~\bibnamefont {Stehr}},\
  }\href {\doibase 10.1016/0550-3213(80)90055-3} {\bibfield  {journal}
  {\bibinfo  {journal} {Nucl.Phys.}\ }\textbf {\bibinfo {volume} {B175}},\
  \bibinfo {pages} {293} (\bibinfo {year} {1980})}\BibitemShut {NoStop}%
\end{thebibliography}%

\end{document}